\documentclass[fleqn,usenatbib]{mnras}

\usepackage[T1]{fontenc}

\DeclareRobustCommand{\VAN}[3]{#2}
\let\VANthebibliography\thebibliography
\def\thebibliography{\DeclareRobustCommand{\VAN}[3]{##3}\VANthebibliography}

\DeclareSymbolFont{matha}{OML}{txmi}{m}{it}% txfonts
\DeclareMathSymbol{\vv}{\mathord}{matha}{29}

\newcommand{\HI}{H{\small{I}}}

\usepackage{graphicx}	% Including figure files
\usepackage{amsmath}	% Advanced maths commands
\usepackage{amssymb}	% Extra maths symbols
\usepackage{comment}
\usepackage{braket}
\usepackage{booktabs}
\usepackage{color}
\title[\HI{} galaxy properties at $z\sim0.37$ in the LSS]{MIGHTEE-H {\textsc i}: H {\textsc i} galaxy properties in the large scale structure environment at $z\sim 0.37$ from a stacking experiment}

\author[F. Sinigaglia et al.]{\hspace{-0.15cm}
Francesco Sinigaglia,$^{1,2,3,4}$\thanks{Email: francesco.sinigaglia@phd.unipd.it}
Giulia Rodighiero,$^{1,2}$
Ed Elson,$^{5}$ 
Alessandro Bianchetti,$^{1,2}$
Mattia Vaccari,$^{6,7,8}$
\newauthor
Natasha Maddox,$^{9,10}$
Anastasia A. Ponomareva,$^{11}$
Bradley S. Frank,$^{6,7,12,13}$
Matt J. Jarvis,$^{11,5}$
\newauthor
Barbara Catinella,$^{14,15}$ 
Luca Cortese,$^{14,15}$ 
Sambit Roychowdhury,$^{16}$
Maarten Baes,$^{17}$
\newauthor
Jordan D. Collier,$^{7,18,19}$
Olivier Ilbert,$^{21}$
Ali A. Khostovan,$^{22}$
Sushma Kurapati,$^{12}$
\newauthor
Hengxing Pan,$^{11}$
Isabella Prandoni,$^{8}$
Sambatriniaina H. A. Rajohnson,$^{12}$
Mara Salvato,$^{23}$
\newauthor
Srikrishna Sekhar, $^{24,7}$
Gauri Sharma$^{5}$
\vspace{0.5cm}
\\
$^{1}$Department of Physics and Astronomy, Università degli Studi di Padova, Vicolo dell’Osservatorio 3, I-35122, Padova, Italy\\
$^{2}$INAF - Osservatorio Astronomico di Padova, Vicolo dell’Osservatorio 5, I-35122, Padova, Italy\\
$^{3}$Instituto de Astrof\'isica de Canarias, Calle Via L\'actea s/n, E-38205, La  Laguna, Tenerife, Spain\\
$^{4}$Departamento  de  Astrof\'isica, Universidad de La Laguna,  E-38206, La Laguna, Tenerife, Spain\\
$^{5}$Department of Physics and Astronomy, University of the Western Cape, Robert Sobukwe Rd, 7535 Bellville, Cape Town, South Africa\\
$^{6}$Inter-university Institute for Data Intensive Astronomy, Department of Physics and Astronomy, University of the Western Cape,\\ 7535 Bellville, Cape Town, South Africa\\
$^{7}$Inter-university Institute for Data Intensive Astronomy, Department of Astronomy, University of Cape Town,\\ 7701 Rondebosch, Cape Town, South Africa\\
$^{8}$INAF - Istituto di Radioastronomia, via Gobetti 101, 40129 Bologna, Italy\\
$^{9}$School of Physics, H.H. Wills Physics Laboratory, Tyndall Avenue, University of Bristol, Bristol, BS8 1TL, UK\\
$^{10}$Faculty of Physics, Ludwig-Maximilians-Universität, Scheinerstr. 1, 81679 Munich, Germany\\
$^{11}$Oxford Astrophysics, Denys Wilkinson Building, University of Oxford, Keble Rd, Oxford, OX1 3RH, UK\\
$^{12}$Department of Astronomy, University of Cape Town, Private Bag X3, Rondebosch 7701, South Africa\\
$^{13}$South African Radio Astronomy Observatory, 2 Fir Street, Black River Park, Observatory, 7925, South Africa\\
$^{14}$International Centre for Radio Astronomy Research (ICRAR), University of Western Australia, 35 Stirling Highway, Crawley,\\ WA 6009, Australia\\
$^{15}$ARC Centre of Excellence for All-Sky Astrophysics in 3 Dimensions (ASTRO 3D), Australia\\
$^{16}$University Observatory, Faculty of Physics, Ludwig-Maximilians-Universität, Scheinerstr. 1, 81679 München, Germany\\
$^{17}$Sterrenkundig Observatorium, Universiteit Gent, Krijgslaan 281 S9, 9000 Gent, Belgium\\
$^{18}$School of Science, Western Sydney University, Locked Bag 1797, Penrith, NSW 2751, Australia\\
$^{19}$CSIRO, Space and Astronomy, PO Box 1130, Bentley, WA, 6102, Australia\\
$^{20}$Centre for Radio Astronomy Techniques and Technologies, Department of Physics and Electronics, Rhodes University, \\ PO Box 94, Makhanda 6140, South Africa\\
$^{21}$Aix Marseille Univ, CNRS, CNES, LAM, Marseille, France\\
$^{22}$Laboratory for Multiwavelength Astrophysics, School of Physics and Astronomy, Rochester Institute of Technology,\\ 84 Lomb Memorial Drive, Rochester, NY 14623, USA\\
$^{23}$Max Planck Institute for Extraterrestrial Physics, Giessembachstrasse 1, D-857498, Garching, Germany\\
$^{24}$National Radio Astronomy Observatory, 1003 Lopezville Road, Socorro, NM 87801, USA\\
}

\date{Accepted XXX. Received YYY; in original form ZZZ}

\pubyear{2023}

\begin{document}
\label{firstpage}
\pagerange{\pageref{firstpage}--\pageref{lastpage}}
\maketitle

\clearpage

%\begin{minipage}[t][-25cm]{0,5\textwidth}
%\begin{flushright}
\begin{abstract}
We present the first measurement of \HI{} mass of star-forming galaxies in different large scale structure environments from a blind survey at $z\sim 0.37$. In particular, we carry out a spectral line stacking analysis considering $2875$ spectra of colour-selected star-forming galaxies undetected in \HI{} at $0.23 < z < 0.49$  in the COSMOS field, extracted from the MIGHTEE-\HI{} Early Science datacubes, acquired with the MeerKAT radio telescope.
We stack galaxies belonging to different subsamples depending on three different definitions of large scale structure environment: local galaxy overdensity, position inside the host dark matter halo (central, satellite, or isolated), and cosmic web type (field, filament, or knot). We first stack the full star-forming galaxy sample and find a robust \HI{} detection yielding an average galaxy \HI{} mass of $M_{\rm HI}=(8.12\pm 0.75)\times 10^9\, {\rm M}_\odot$ at $\sim 11.8\sigma$. Next, we investigate the different subsamples finding a negligible difference in $M_{\rm HI}$ as a function of the galaxy overdensity. We report an \HI{} excess compared to the full sample in satellite galaxies ($M_{\rm HI}=(11.31\pm1.22)\times 10^9$, at $\sim 10.2 \sigma$) and in filaments ($M_{\rm HI}=(11.62\pm 0.90)\times 10^9$. Conversely, we report non-detections for the central and knot galaxies subsamples, which appear to be \HI{}-deficient. We find the same qualitative results also when stacking in units of \HI{} fraction ($f_{\rm HI}$).
We conclude that the \HI{} amount in star-forming galaxies at the studied redshifts correlates with the large scale structure environment.
\end{abstract}
%\end{flushright}
%\end{minipage}

\vspace{1cm}
%\clearpage

%%%%%%%%%%%%%%%%%%%%%%%%%%%%%%%%%%%%%%%%%%%%%%%%%%

%%%%%%%%%%%%%%%%% BODY OF PAPER %%%%%%%%%%%%%%%%%%

\section{Introduction}

%The \HI{} is important in galaxy properties. Galaxy properties correlate with large-scale environment. The large-scale cosmic web environment emerges as growth of cosmic structures. We are interested in studying the correlation between HI properties and LSS with a spectral stacking experiment. 

The evolution of galaxies, including mass assembly, star formation and morphological transformations of galaxies, is known to be strongly connected to the availability of fresh molecular Hydrogen (H$_2$) supporting the star formation process. Star-forming H$_2$ clumps arise via gravitational instability and collapse out of large diffuse \HI{} clouds, whose existence represents therefore the necessary condition to trigger the formation of new stars. While part of the galactic \HI{} can be produced by either recombination of ionized Hydrogen (H{\small II}) or other reprocessing mechanisms of the internal gas, hydrogen accretion on galaxies from the circumgalactic and intergalactic media plays a fundamental role to ensure the availability of \HI{} reservoirs.

In this context, the environment surrounding a galaxy and its physical conditions may assume a primary importance in regulating gas accretion and removal, galaxy interactions, and other relevant evolutionary phenomena.
In fact, galaxy surveys have measured the position of millions of galaxies and demonstrated that at cosmological scales smaller than a few hundred Mpc, matter is no longer uniformly distributed, but forms a filamentary pattern called \textit{cosmic web} \citep[e.g.][]{Bond1996}, which is constituted by filaments interconnecting massive knots and surrounding large voids. In the standard inflationary $\Lambda$CDM scenario, the cosmic web naturally emerges from gravitational instability and growth of cosmic structures \citep[e.g.][]{Zeldovich1970,Bond1996}, happening as a result of the presence of small inhomogeneities in the primordial matter field. While driven by dark matter, the formation of the cosmic web also involves baryons. As a result, gas exhibits different properties (temperature, rotation and dispersion velocities, ionization state, among others) depending on the environment it lives in \citep[e.g.][and references therein]{Martizzi2019,Galarraga2020,Sinigaglia2021}, and therefore it may impact the evolution of the galaxies it feeds in a different way. The large scale structure has been shown to correlate with several galaxy properties, for instance star formation history and quenching \citep[e.g.][and references therein]%{Darwish2014,Ricciardelli2014,Snedden2016,Vulcani2019,Alam2019,Salerno2019,Kraljic2019,Kraljic2020,Malavasi2022}
{Darwish2014,Vulcani2019,Kraljic2020,Malavasi2022}, colour \citep[e.g.][]{Chen2017,Pandey2020}, stellar mass and its assembly \citep[e.g.][]{Alpaslan2016,Chen2017,Malavasi2017,Kraljic2020}, angular momentum magnitude and alignment \citep[e.g.][]{Libeskind2012,Tempel2013,Krowleski2019,Barsanti2022}, and the stellar mass - gas metallicity relation \citep{Donnan2022}, among others.   

In this picture, the role of \HI{} has been investigated only in the nearby universe at $z<0.1$ \citep[e.g.][]{Kleiner2017,COdekon2018,Tudorache2022,Cortese2021}, while it has remained unexplored at higher redshift due to the difficulty in detecting the 21-cm emission line and the general lack of deep complete spectroscopic surveys covering volumes large enough to reconstruct the cosmological large-scale environment. 

Statistical approaches such as spectral line stacking \citep{Zwaan2001} can be adopted to exploit the property of Gaussianity of the noise and to extract a global mean \HI{} signal out of the investigated population, at the expense of the information about the \HI{} content of the individual galaxies constituting the sample. The spectral line stacking technique has been widely used in the last decades to probe the \HI{} content in galaxies at different redshift, and in particular to unveil correlations such as the presence and abundance of \HI{} in galaxy clusters \citep{Zwaan2000,Chengalur2001,Lah2009,Healy2021}, \HI{} galaxy scaling relations \citep{Fabello2011b,Gereb2015,Brown2017,Sinigaglia2022a,Bera2022,Chowdhury2022,Bera2023,Pan2023}, the \HI{} mass function \citep{Pan2020,Bera2022}, the $M_{\rm HI}$ content of AGN host galaxies \citep{Fabello2011b,Gereb2013,Gereb2015}, the baryonic Tully-Fisher relation \citep{Meyer2016}, the \HI{} cosmic density evolution with redshift \citep{Lah2007,Delhaize2013,Kanekar2016,Rhee2018,Bera2019,Chowdhury2020,Chen2021,Chowdhury2021,Chowdhury2022a,Chowdhury2022b,Chowdhury2022}, the \HI{} content of galaxy groups and the $M_{\rm HI}$-$M_{\rm halo}$ relation \citep[][]{Guo2020,Chauhan2021,Roychowdhury2022,Dev2023}, among others.  Spectral line stacking has been successfully applied to other spectral lines as well, such as CO and CII lines \citep[e.g.][]{Decarli2018,Bischetti2019,Jolly2021,Romano2022}. 

The problem of the lack of a deep complete spectroscopic coverage on large sky areas can be alleviated in the case where multi-wavelength photometric observations can provide an accurate photometric redshift estimation. In this case, photometric redshifts can be used to perform tomographic analysis on the sky in thick redshift slices. In the COSMOS field -- studied in this work -- the exceptional amount of work over the past decades has made it possible to collect an incredibly large dataset, with photometry ranging from the X-ray to the radio domain  \citep[see e.g.][]{Laigle2016,Weaver2022}. 

In this work we perform a systematic investigation of the correlations between the content of \HI{} in galaxies and different definitions of the large scale structure environment at a median redshift $z\sim 0.37$, then compare to reference results at $z\sim 0$. %, to whether such correlations can be responsible of driving the observed trends in the galaxy evolution picture. 
We perform a \HI{} stacking experiment %(and in particular, for the first time at such high redshift)
and measure the amount of \HI{} in different environments, based on subsamples defined from galaxies located in low-/high-density environments, in field/filaments/knots, or being central/satellite/isolated inside their host dark matter halo.

This paper is organized as follows. In \S\ref{sec:mightee} we present the details of the MIGHTEE survey and of the \HI{} data we make use of. In \S\ref{sec:lss} we summarize the large-scale computations and definitions used throughout the work. \S\ref{sec:sample} defines the global galaxy sample we analyze and \S\ref{sec:stacking} introduces our stacking procedure. In \S\ref{sec:results} we present the results we obtain and in \S\ref{sec:discussion} we provide a discussion and contextualization of them. We conclude in \S\ref{sec:conclusions}. 

Throughout the paper, we assume a cosmology as reported from the most recent results from the Planck satellite, with $\Omega_{\rm m}=0.311$, $h=0.677$, $\Omega_k=0$ \citep[TT,TE,EE+lowE+lensing+BAO,][]{Planck2018}.

\section{H{\small I} data from MIGHTEE} \label{sec:mightee}

MeerKAT is the SKA precursor located in South Africa and comprises 64 offset Gregorian dishes ($13.5$ m diameter main reflector and $3.8$ m sub-reflector), equipped with receivers in UHF–band ($580< \nu <1015$ MHz), L–band ($900< \nu <1670$ MHz), and S–band ($1750< \nu <3500$ MHz).

The MeerKAT International GigaHertz Tiered Extragalactic Exploration Large Survey Program \citep[MIGHTEE,][]{Jarvis2016} is a survey, conducted with the MeerKAT radio interferometer \citep[e.g. ][]{Jonas2016}%,Camilo2018,Mauch2020}
. MIGHTEE is targeting at L- and S-bands four deep, extragalactic fields (COSMOS, XMM-LSS, CDFS, ELAIS-S1), characterized by a wealth of multi-wavelength data made available by past and ongoing observational efforts. %currently being observed by MeerKAT, the SKA precursor  located in South Africa (Jonas 2009). 

The MIGHTEE data are acquired in spectral and full Stokes mode, thereby making MIGHTEE a spectral line, continuum and polarization survey. In this paper, we make use of the Early Science \HI{} spectral line data from MIGHTEE, presented in  \citet[][]{Maddox2021}.
The observations were conducted between mid-2018 and mid-2019 and targeted $\sim3.5$ deg$^2$ in the XMM-LSS field and $\sim1.5$ deg$^2$ in the COSMOS field. These observations were performed with the full array (64 dishes) in L–band, using the 4k correlator mode ($209\,\rm{kHz}$, corresponding to $52$ km s$^{-1}$ at $z=0.23$ and $56$ km s$^{-1}$ at $z=0.49$). Our analysis is limited to the redshift interval $0.23<z<0.49$, excluding the spectral bands covering $0.09<z<0.23$ and $z>0.49$ from our analysis because of bad radio frequency interference (RFI) contamination. We use the MIGHTEE-\HI{} data covering only the COSMOS field data because they are the only data currently available in the redshift range of interest ($0.23<z<0.49$) as part of the MIGHTEE Early Science data release. At the investigated redshifts, MIGHTEE-\HI{} data have noise with well-behaved Gaussian properties. The median
\HI{} noise rms increases with decreasing frequency, ranging from $85 \, \mu{\rm Jy} \, {\rm beam}^{-1}$ at $\nu\sim 1050$ MHz to $135 \, \mu{\rm Jy}\, {\rm beam}^{-1}$ at
$\nu \sim 950$ MHz.

The MIGHTEE–\HI{} Early Science visibilities were processed with the \texttt{ProcessMeerKAT} calibration pipeline\footnote{\url{https://idia-pipelines.github.io/docs/processMeerKAT}}. The pipeline is \texttt{CASA}-based and performs standard calibration routines and strategies for the spectral line data such as flagging, bandpass and complex gain calibration. The continuum subtraction was done in both the visibility and image domain using standard \texttt{CASA} routines \textsc{uvsub} and \textsc{uvcontsub}. Residual visibilities after continuum subtraction were imaged using \texttt{CASA}’s task \textsc{tclean} with Briggs' weighting (\textsc{robust}=0.5). Median filtering was applied to the resulting datacubes to reduce the impact of errors due to continuum subtraction. %A full description of the data reduction strategy and data quality assessment will be presented in Frank et al. (in prep).

\begin{table}
    \centering
    \begin{tabular}{ll}
    \toprule
    \toprule
    \textbf{MIGHTEE-\HI{} data} & \\
    %\toprule
    \toprule
    \textbf{Survey parameter} & \textbf{Value}\\
    %\midrule
    \midrule
    Field & COSMOS \\
    Area     & $1.5$ deg$^2$ \\
    Integration time & $16$h \\
    Frequency resolution & $209$ kHz\\
    Recession velocity resolution & $52$ km s$^{-1}$ at $z=0.23$\\
    Frequency range & $0.950-1.050$ GHz \\
    Recession velocity range & $68952-146898$ km s$^{-1}$ \\
    Beam (FWHM) &  $14.5^{\prime\prime}\times11.0^{\prime\prime}$ \\
    %Pixel size     & $2.0^{\prime\prime}$ \\
    %Image size & $4096\times4096$ pixels \\
    %Median HI channel rms noise & $XYZ$ $\mu$Jy beam$^{-1}$\\
    %$3\sigma$ HI column density sensitivity & $XYZ$ cm$^{-2}$
%(per channel)\\
    \bottomrule
    \bottomrule
    \end{tabular}
    \caption{Summary of the details of MIGHTEE-\HI{} data presented in \S\ref{sec:mightee} and used in this paper.}
    \label{tab:mightee}
\end{table}

% *********************************************
%**********************************************
% *********************************************

% *********************************************
%**********************************************
% *********************************************
\section{large scale structure definition} \label{sec:lss}

In this section we summarize the procedure used to construct the density field and to compute the large-scale scale properties in the COSMOS field, as derived by \cite{Darwish2015b}. For more details, we refer the reader to \cite{Darwish2015b}.

\subsection{Construction of the density field}

In absence of a complete spectroscopic coverage over the full $\sim2$ deg$^2$ of the COSMOS field, the density field is constructed from the photometric COSMOS2015 catalogue \citep{Laigle2016}. The redshift uncertainty associated with photometric redshifts (photo-$z$s hereafter) is too large to accurately unveil redshift-space distorsions effects. Nonetheless, an accurate photo-$z$ determination ($\Delta z/(1+z)\lesssim 0.01$) can still enable a robust density field construction, sufficient to study statistical relations between galaxy properties and the surrounding large-scale environment. Following \cite{Darwish2015b}, a cut in $M_*$ is applied at $\log_{10}(M_*/{\rm M}_\odot)\gtrsim 9.6$, to ensure that the photometric sample is $90 \%$ complete in $M_*$. This guarantees that our sample is volume-limited, and hence, we do not expect biases arising from an evolving mass cut across redshift. We will use this fact later in the paper. 

The photo-$z$ estimation for the COSMOS2015 catalogue was performed by means of the spectral energy density fitting technique, over $>30$ photometric bands from the near-UV to the near-IR domains. When compared to the spectroscopic sample from the zCOSMOS survey \citep{Lilly2009}, the photometric COSMOS sample turns out to have a median accuracy $\Delta z/(1+z)_{\rm median}\sim 0.007$, and to have accuracy $\Delta z/(1+z)\lesssim 0.01$ out to $z\sim 1.2$ (see Fig. 1 in \citet{Darwish2015b}). 

The density field is constructed in adjacent redshift slices of variable width, and each galaxy is associated with a weight corresponding to the fraction of the photo-$z$ probability distribution of the galaxy lying inside each $z$-slice, i.e. the likelihood of a galaxy having redshift included in that redshift interval. A galaxy is used to construct the density field only in the $z$-slices where their weight is $\ge 10\%$.  

After having assigned weights to galaxies, \cite{Darwish2015b} employed the weighted adaptative kernel smoothing technique (with weights corresponding to the ones discussed above) to estimate the density field, employing a 2D Gaussian kernel with variable variance depending on the local galaxy density. The adopted global smoothing scale is $s=0.5\, \rm{Mpc}$, roughly the virial radius of galaxy clusters in the COSMOS field \citep[e.g.][]{Finoguenov2007}.

The density field is estimated via interpolation at the galaxy positions, using RA and DEC as angular coordinates and the maximum-likelihood redshift slice as spectral coordinate.

\begin{figure}
    \centering
    \includegraphics[width=\columnwidth]{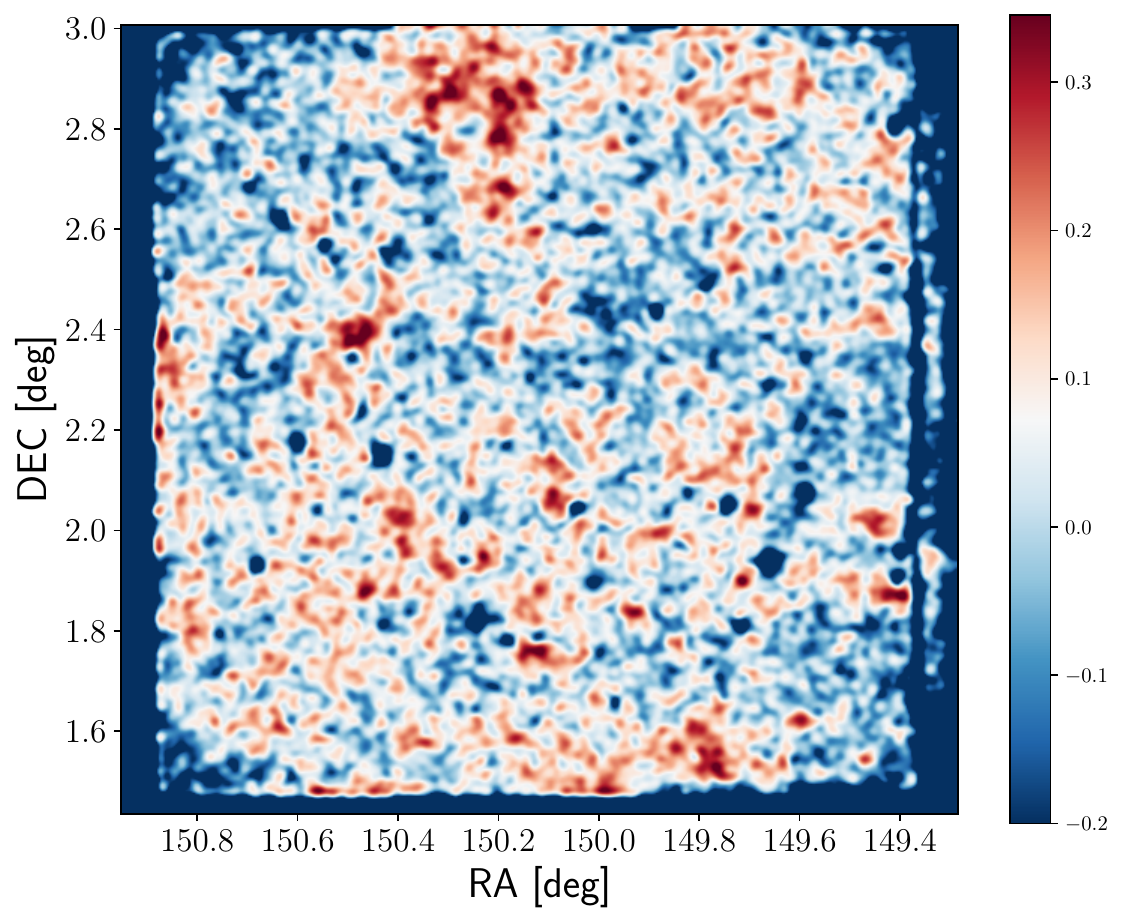}
    \caption{Projected density field, colour-coded from blue to red at increasing density. The plot has been generated by first binning galaxy positions (RA and DEC) of all galaxies in the catalogue with photometric redshift $0.1<z_{\rm phot}<0.6$ in pixels of size $d\sim 10^{\prime\prime}$, then applying a two-dimensional Gaussian smoothing with standard deviation $\sigma=30^{\prime\prime}$, and eventually applying a bilinear interpolation. The colorbar is reported in units of $\log_{10}(1 + n_{\rm gal})$, where $n_{\rm gal}$ is the galaxy number counts per pixel.
    One can visually identify overdense regions with spherical or ellipsoidal shapes (knots) and with elongated one-dimensional shapes (filaments), as well as underdense regions (field).}   
    \label{fig:densfield}
\end{figure}

Fig.~\ref{fig:densfield} shows the projected density field, colour-coded from blue to red at increasing density. Even though we perform analysis at $0.23<z_{\rm spec}<0.49$, in this plot we consider $0.1<z_{\rm phot}<0.6$, i.e. a larger redshift interval, to take into account the uncertainty in photometric redshift estimation of the galaxies with $z_{\rm phot}$ at the edges of the spectroscopic redshift interval ($z_{\rm phot}\sim 0.23$ and $z_{\rm phot}\sim 0.49$). One can visually identify overdense regions with spherical or ellipsoidal shapes (knots) and with elongated one-dimensional shapes (filaments), and underdense regions (field).

\begin{figure}
    \centering
    \includegraphics[width=\columnwidth]{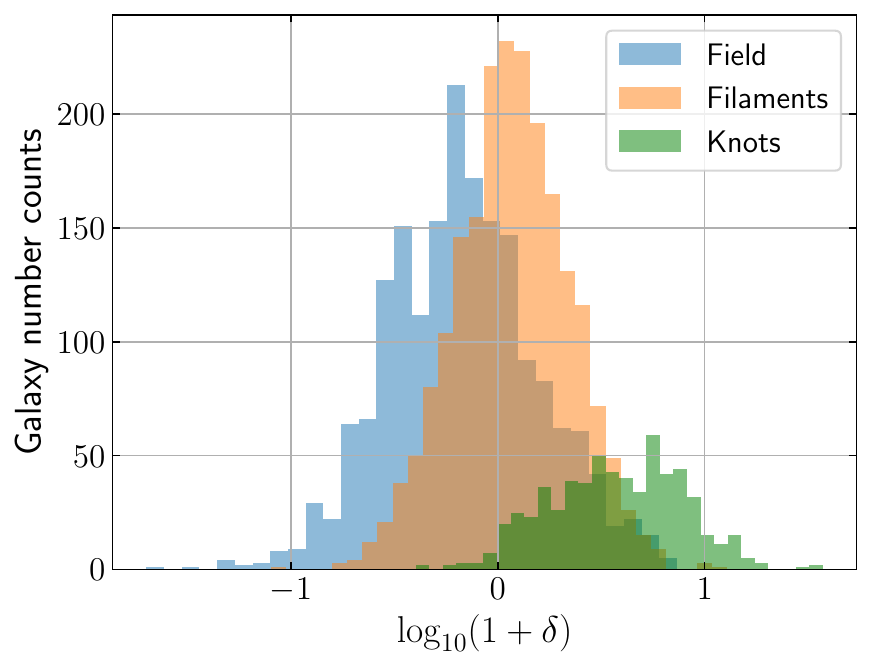}
    \caption{Distributions of field (blue), filaments (orange) and knots (green) galaxies as a function of galaxy overdensity (in units $\log_{10}(1+\delta)$, to visually enhance the differences). As expected, the histograms clearly evidence that field galaxies mostly live in low-density regions, filaments cover the intermediate-density regime, while knots galaxies are preferentially found in high-density regions.}   
    \label{fig:histenvdens}
\end{figure}

\subsection{Cosmic web environments definition} \label{sec:cw_def}

The classification of the large-scale cosmic web environment performed by \cite{Darwish2015b} consists in the 2D version of the Multi-scale Morphology Filter algorithm \citep{AragonCalvo2007,Darwish2014}. 
In particular, this algorithm relies on the computation of the local curvature of the density field and the shape of structures, i.e. on geometrical arguments. The local curvature and shape are computed from the eigenvalues $\lambda_{\rm i}$ of the rank-2 Hessian matrix (with derivatives computed along the angular coordinates) $\mathcal{H}=\partial_{\rm i}\partial_{\rm j}\delta_z(\bf{x})$ of the density field $\delta$, at coordinates $\bf{x}$ and redshift $z$. Defining $\lambda_{1,2}$ such that $|\lambda_2|\ge|\lambda_1|$, one defines cosmic web type masks $\epsilon_{\rm k}$ and $\epsilon_{\rm f}$ at coordinates ${\bf x}$ as:
\begin{itemize}
    \item $\epsilon_{\rm k}=1$ if $\lambda_1\le0$ and $\lambda_2\le0$, $0$ otherwise, for knots;
    \item $\epsilon_{\rm f}=1$ if $\lambda_1\le 0$, $0$ otherwise, for filaments.
\end{itemize}

This classification implies that the point $\bf{x}$ needs to be  a local minimum, or a saddle point, in order to sit in a knot or in a filament. If point $\bf{x}$ fulfills both the aforementioned $\epsilon=1$ conditions, one defines the functions:
\begin{equation}
D_{\rm k}=|\lambda_1|/|\lambda_2|, \, D_{\rm f}= 1 - D_{\rm k}
\end{equation}
to evaluate the local resemblance of an identified structure to a knot and to a filament, respectively. This definition implies that, if $|\lambda_1|\sim|\lambda_2|$, the slope of the density field has a rotationally-symmetric structure and the point belongs to a knot ($D_{\rm k}\sim 1$). Instead, if $|\lambda_1|\ll|\lambda_2|$, a structure is elongated in one dimension, and is identified as a filament ($D_{\rm f}\sim 1$). The Hessian matrix of the density field has been used as an alternative to the widely-used cosmic web classification based on the Hessian matrix of the gravitational tidal tensor and has been shown to be important to parametrize the properties of large scale structure \citep[see e.g.][]{HeavensPeacock1988,Sinigaglia2021,Sinigaglia2022c}.   

Subsequently, to control the selection of knots and filaments, the functions $D_{\rm f}$ and $D_{\rm k}$ are non-linearly transformed into $M_{\rm k}$ and $M_{\rm f}$, respectively, as: 
\begin{equation}
    M_{\rm k} = \exp\left(-\frac{D_{\rm f}}{\beta^2}\right)
\end{equation}
\begin{equation}
    M_{\rm f} = \exp\left(-\frac{D_{\rm k}}{\beta^2}\right)
\end{equation}
where $\beta$ is a parameter which regulates the criterion used to select features, set to $\beta=0.5$ in \cite{Darwish2015b}. This transform has the effect of enhancing the differences between $D_k$ and $D_f$, trying to discriminate more clearly between different cosmic web environments. The quantities $M_k$ and $M_f$ should then be seen as boosted knot and filament signals respectively, and represent auxiliary variables to be used later on in the computations.

Eventually, to maximize the significance of the extracted large scale structure features and to best separate between such structures and the background, one can not only exploit the signs and the ratios between eigenvalues, but also their magnitude. In fact, the magnitude of eigenvalues will tend to be small for the background due to its fluctuating nature, while it will tend to be large for real structures. 
This fact can be accounted for by defining the norm  of the Hessian as
\begin{equation}
    I = 1- \frac{\sqrt{\lambda_1^2+\lambda_2^2}}{2\,c^2}
\end{equation}
where $c=0.5\times {\rm max}(\sqrt{\lambda_1^2+\lambda_2^2})$ at each $z$-slice \citep{Frangi1998}. 

The knot and filament probability signals are then evaluated for each pixel on the set of maps obtained using different physical scales $L$ as $S_L=\epsilon \times M \times I$, and eventually the signal $S$ is chosen to be the maximum among the different scales, i.e. $S={\rm max}_L(S_L)$.

At this point, a knot and filament signal ($S_k$ and $S_f$ respectively) are associated to each galaxy. To conclusively establish what environment a galaxy belongs to, \cite{Darwish2015b} found the optimal signal cuts to be parametrized by $t_k=0.0639\times z+0.1142$ and $t_f=0.0253\times z+0.0035$ for knots and filaments, as the best trade-off that guarantees to have a sufficiently large sample and to minimize contamination. 

Eventually, a galaxy is classified to belong to a knot, a filament, or to the field as follows:
\begin{itemize}
    \item knot, if $S_k>t_k$ and $S_f<t_f$, or $S_k>t_k$ and $S_f>t_f$ and $S_k>S_f$;
    \item filament, if $S_k<t_k$ and $S_f>t_f$, or $S_k>t_k$ and $S_f>t_f$ and $S_k<S_f$;
    \item field, otherwise.
\end{itemize}

To better understand which density regimes the three classes of cosmic web environment correspond to, Fig.~\ref{fig:histenvdens} shows the distributions of field (blue), filaments (orange) and knots (green) galaxies as a function of galaxy overdensity\footnote{We notice that \cite{Darwish2015b} denote with 'overdensity' $1+\delta$, with $\delta=\rho_{\rm g}/\bar{\rho}_{\rm g}-1$ and $\rho_{\rm g}$ and $\bar{\rho}_{\rm g}$ the galaxy density and mean galaxy density respectively. For consistency with cosmological studies we refer to only $\delta$  as 'overdensity'.} (in units $\log_{10}(1+\delta)$, to visually enhance the differences). As expected, the histograms clearly evidence that field galaxies mostly live in low-density regions, filaments cover the intermediate-density regime, while knots galaxies are preferentially found in high-density regions.

\subsection{Central, satellite, and isolated galaxies}

To define whether each galaxy in the analyzed sample is a central or satellite in a dark matter halo, \cite{Darwish2017} first define a catalogue of galaxy groups and afterwards flag as central the most massive galaxy in a group and as satellite the remaining galaxies of the group. Isolated galaxies, which are found not to be associated with any group, have two possible interpretations. One is that they are central galaxies with too faint satellites to be detected. Alternatively, they can be regarded as satellite galaxies which have been ejected from their group, or as the product of an earlier merger of a galaxy pair.

Galaxy groups are determined via the application of a friends-of-friends algorithm \citep{HucraGeller1982}, based on the ansatz that any pair of galaxies closer than a given critical distance (also called \textit{linking length}) belongs to the same group. The definition of the linking length is therefore of primary importance to obtain a reliable galaxy group classification. The angular and redshift separation between two galaxies $i$ and $j$ are defined by \cite{Darwish2017} as
\begin{equation}
    \Delta\theta_{ij}\le b_{\rm ang}\, D_c(z)^{-1}\, n(z)^{-1/2} \, , \, |z_i-z_j| \le 1/\left(b_z\, \sigma_{\Delta z/(1+z)}\right) 
\end{equation}
where $D_c(z)$ is the comoving distance and $n(z)$ is the median number density of galaxies at redshift $z$, and $b_{\rm ang}$ and $b_z$ are free parameters set to $b_{\rm ang}=1.3$ and $b_z=1.5$. We refer to \cite{Darwish2017} for a thorough discussion of the results. 

% *********************************************

\begin{figure*}
    \centering
    \includegraphics[width=\textwidth]{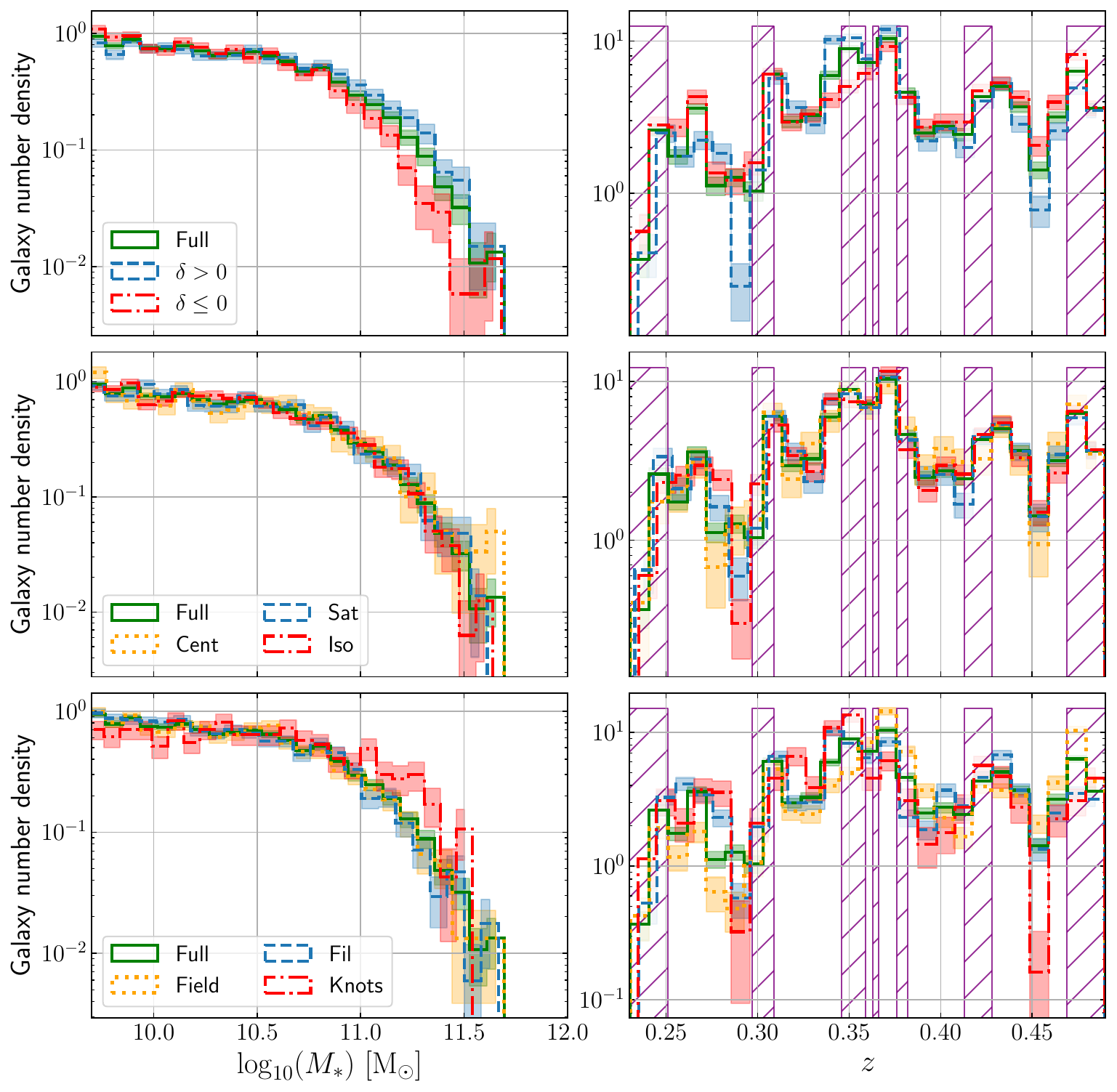}
    \caption{$\log_{10}(M_*)$ (left) and $z$ (right) distributions of the full sample (green solid) and different subsamples investigated in this work. Top row: high-density $\delta>0$ (blue dashed) and low-density $\delta\le 0$ (red dashed-dotted) galaxies. Mid row: central (yellow dotted), satellite (blue dashed), and isolated (red dashed-dotted) galaxies. Bottom row: field (yellow dotted), filaments (blue dashed), and knots (red dashed-dotted) galaxies. Poisson uncertainties are shown as shaded areas. In the right column, purple hatched areas indicate frequency intervals strongly affected by RFI, for which we adopt the RFI flagging and masking described in \S\ref{sec:stacking}.}
    \label{fig:prop}
\end{figure*}

\begin{figure*}
    \centering
    \includegraphics[width=\textwidth]{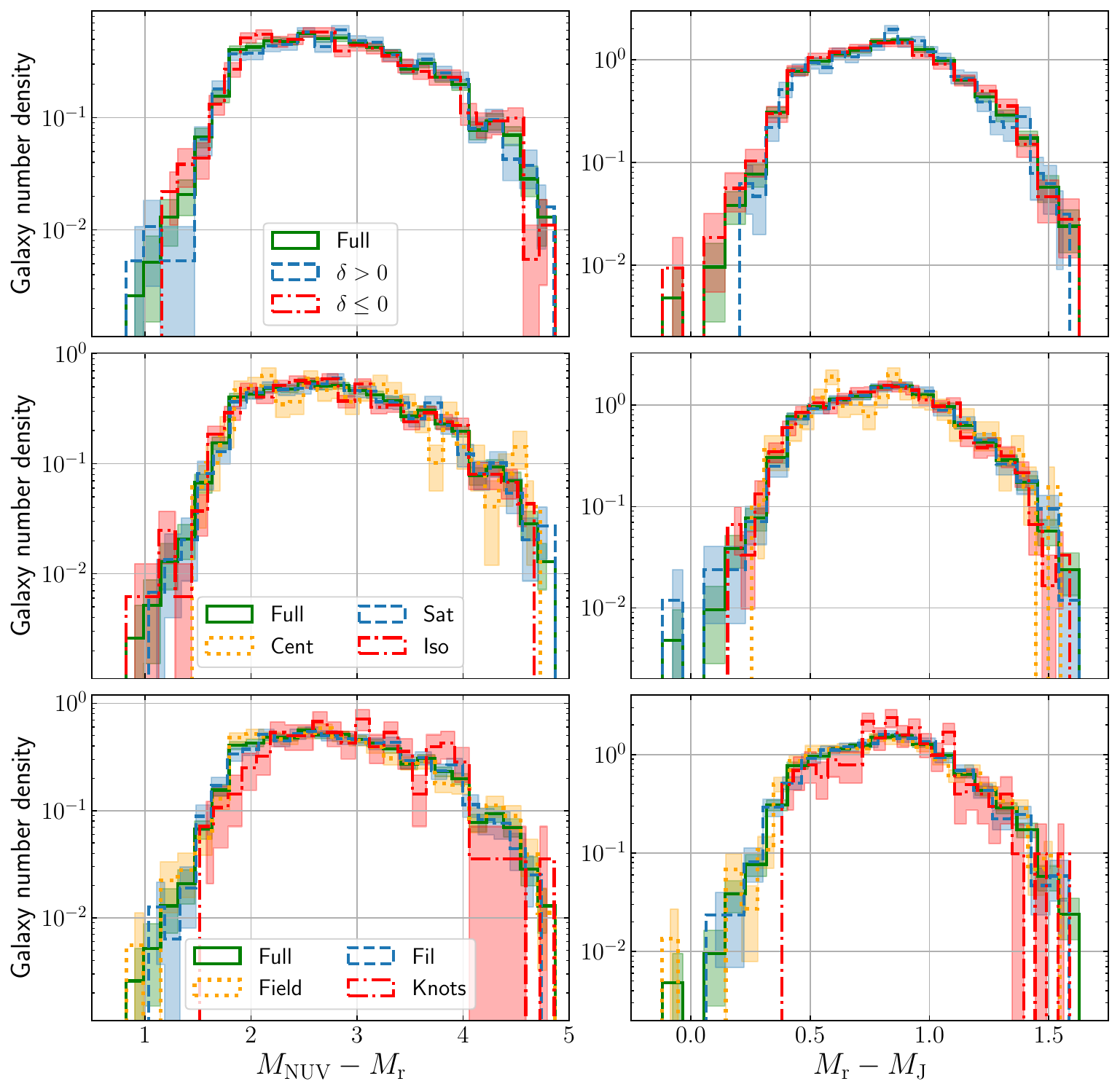}
    \caption{Same as Fig. \ref{fig:prop}, but for rest-frame colours $M_{\rm NUV}-M_{\rm r}$ (left) and $M_{\rm r}-M_{\rm J}$ (right).}
    \label{fig:prop_colour}
\end{figure*}

\section{Sample selection}\label{sec:sample}

In this section we present the details about the sample selection.

\subsection{Cross-match with spectroscopic catalogues}

The galaxy sample we make use of in this work consists of star-forming galaxies at redshift $0.23<z<0.49$ in the COSMOS field \citep{Scoville2007}, with spectroscopic redshift available from public surveys. Compared to the one used in \cite{Sinigaglia2022a}, the catalogue has been updated with the addition of new sources. In particular, we cross-match all the galaxies contained in the catalogue compiled by \cite{Darwish2015b,Darwish2017} (including both star-forming and passive galaxies) with the list of spectroscopic redshifts compiled by querying publicly-available catalogues from multiple surveys of the COSMOS field (Khostovan et al., in prep.), and keep only those galaxies for which we were able to identify a spectroscopic counterpart. As shown in \cite{Sinigaglia2022a}, the spectroscopic counterpart sample of the COSMOS photometric catalogue is complete (in terms of representativity of the photometric sample) down to $\log_{10}(M_*)\sim9.5$, roughly the same mass limit used to define the catalogue used in this work. At this point, we select only star-forming galaxies from the resulting sample, based on a rest-frame colour-colour $NUV-r/r-J$ plane selection, extracted directly from the \cite{Laigle2016} catalogue\footnote{We notice that, even though a more recent COSMOS photometric catalogue has been made publicily available \citep{Weaver2022}, we prefer to stick to the \cite{Laigle2016} catalogue for consistency with the computations performed by \cite{Darwish2015b,Darwish2017}. Also, at the probed redshift, the gain in NIR photometry -- which is the main progress between the \cite{Laigle2016} and the \cite{Weaver2022} catalogues -- is not significant. What drives the accuracy at this redshift are the medium bands, and they did not change from one catalogue to the other.}. In particular, according to this selection, quiescent galaxies are defined as those featuring $M_{NUV}-M_r > 3(M_r - M_J) + 1$ and $M_{NUV} - M_r > 3.1$, while the remaining galaxies are regarded as star-forming.  

After applying these cuts, the resulting catalogue is found to contain $1835$ more galaxies with respect to the catalogue employed in \citet{Sinigaglia2022a}, uniformly spread over the probed redshift range ($0.23<z<0.49$).

Through a further cross-match with the radio-selected AGN catalogue built by \citet{Smolcic2017}, we estimate the fraction of AGNs in our sample to be $\sim 3.5\%$. We do not exclude AGNs from our sample, and their impact on results from \HI{} spectral stacking will be addressed in future publications of the MIGHTEE collaboration. Eventually, by relying on the spectroscopic redshifts we also exclude from our sample all galaxies whose \HI{} emission is expected at radio frequencies strongly affected by RFI. While we cross-checked that RFI does not have a major impact on our stacking results, we prefer to limit our sample to RFI-free regions to have a higher degree of control of systematics.

After the application of all the selection criteria described above, our final sample consists of $2875$ galaxies, with a median redshift $z\sim0.37$. 

Fig.~\ref{fig:prop} illustrates the $\log_{10}(M_*)$ (left)
%, $\log_{10}({\rm SFR})$ (centre) 
and $z$ (right) distributions of galaxies, for different values of the local overdensity field $\delta$ ($\delta \le0$, $\delta>0$; first row), position inside the host dark matter halo (central, satellite, or isolated; second row), and the cosmic web environment they belong to (field, filaments, or knots; third row). We also report average and median $M_*$ and $z$ in Table \ref{tab:results}.
By visually inspecting these distributions, one can easily notice that there are similarities and differences between them. To quantify how similar these distribution are, we have run two-samples one-dimensional Kolmogorov-Smirnov tests comparing galaxy properties distributions in different environments. We report the results in Appendix A, showing the resulting $p$-values graphically in Figs. \ref{fig:ks_mass} and \ref{fig:ks_z}. In the case of the KS test comparing the $M_*$ distribution for different subsample, $\delta>0$, centrals and knots galaxies are the subsamples which tend to yield $p<0.05$. In the case of the KS test comparing the $z$ distributions, we can reject the null hypothesis in most of the cases. 

Nonetheless, it may be misleading to conclude that there are selection biases.

In fact, the systematic differences occurring between high-density and low-density environments (both in terms $\delta$ and of cosmic web types) are known effects in the literature. The stellar (luminosity) mass functions of star-forming galaxies in high-density environments feature an excess of probability in the high-mass (bright) end and, consequently, a dearth of probability towards low masses (luminosities) \citep[see e.g.][]{Bolzonella2010,Davidzon2016}. Therefore, the differences between the $M_*$ distributions do not arise as a result of selection biases. Rather, they originate as a natural consequence of the diversity of physical processes shaping the mass assembly history in distinct environment.
To minimize the impact of such systematic differences, we adopt as target quantity of our stacking not only $M_{\rm HI}$, but also the \HI{} fraction $f_{\rm HI}\equiv M_{\rm HI}/M_*$, as will be explained in more details in \S\ref{sec:stacking}.   
 
As for the $z$ distributions, the Kolmogorov-Smirnov tests tell that we can reject the null hyphotesis in most of the cases. However, we also notice that the deviations between the $z$ distributions of the different subsamples do not show systematic trends, but rather random fluctuations.

Fig.~\ref{fig:prop_colour} shows the $M_{\rm NUV}-M_{\rm r}$ (left) and $M_{\rm r}-M_{\rm J}$ (right) colour distributions, with the same scheme as Fig. \ref{fig:prop}. The distributions of different subsamples do not present evident differences, except for a cutoff of the knots galaxies at redder colours towards the bluer end. However, this latter fact is in part due to galaxies in such high overdensities naturally being redder -- a well-known fact from the literature -- but it is also enhanced by the poor statistics of the knots sample and by the resulting larger statistical fluctuations. In any case, it regards only a tiny fraction of the knots galaxies subsample. This means that the comparison we are performing relates subsamples with consistent underlying colour (and hence SFR) distributions.

We therefore conclude that there are no obvious selection biases at this point. 

\subsection{Cosmic variance assessment}

For a robust analysis of correlations between the large scale structure and \HI{} galaxy properties to be robust, we need to address the possible effect of cosmic variance. In fact, the universe starts to be homogeneous on scales $\sim$ hundreds of Mpc. If the volume is not large enough, the results may be biased by the fact that we are looking at a specific cosmic realization.

To address this and to take into account that the surveyed volume extends over a large radial distance, we compute the comoving volume probed by our observations. %Assuming a cosmology as reported from the most recent results from the Planck satellite, with $\Omega_{\rm m}=0.311$ and $h=0.677$ \citep[][]{Planck2018}, 
Given the assumed cosmology, the observed comoving area is $\Delta{\rm RA}\times\Delta{\rm DEC}=24.2\times 20.1 \, {\rm cMpc}^2$ at $z=0.23$ and  $\Delta{\rm RA}\times\Delta{\rm DEC}=48.0\times  40.1 \, {\rm cMpc}^2$ at $z=0.49$, while the total observed comoving radial distance is $d=949.6\, {\rm cMpc}$. Overall, the total volume is $V=1.07\times10^6\,{\rm cMpc}^3$, equivalent to a cubic volume with side $l\sim 102.3\,{\rm cMpc}$. We stress this volume corresponds to the volume spanned by the spectroscopic sample, not by the entire MIGHTEE datacube, the latter being in principle much larger. In fact, existing spectroscopic surveys in the COSMOS field beyond the nearby universe cover mostly the central part of the field, leaving the angular outskirt of the MIGHTEE data unprobed. 

Even though the total volume is not large enough to host outlier rare cosmic structures such as super-clusters or super-voids, it encompasses a volume comparable to state-of-the-art cosmological hydrodynamic simulations. Therefore, we conclude that, while the probed volume is not very large in a cosmological sense and some cosmic variance effects may still be present, this should not constitute a major issue.

\begin{figure}
    \centering
    \includegraphics[width=\columnwidth]{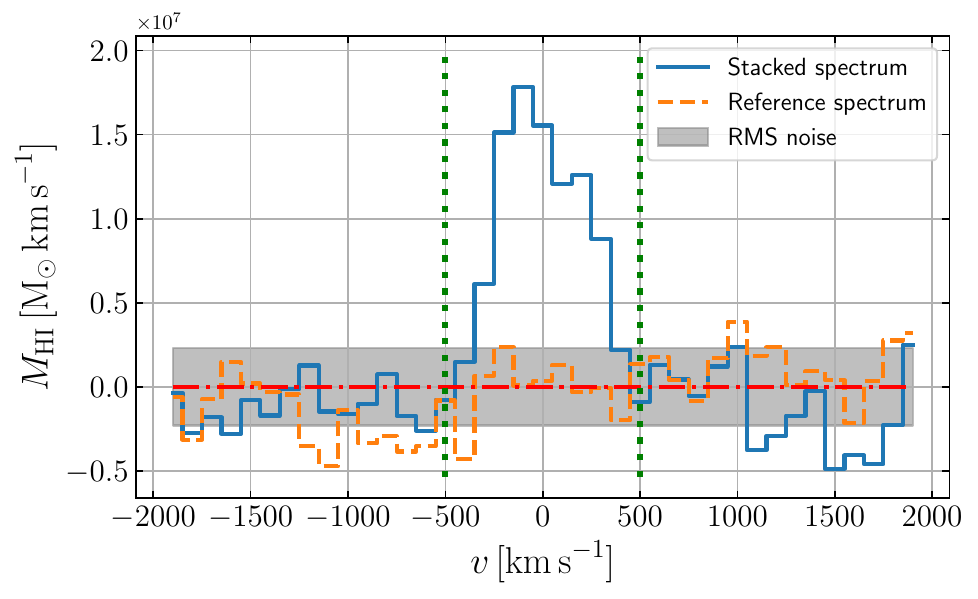}
    \caption{$M_{\rm HI}$ stacked spectrum obtained from the full sample. The blue solid line represents the stacked spectrum, the orange dashed line and the grey shaded area represent the stacked reference spectrum and its rms, respectively, and the green dotted vertical lines represent the mass integration limits.}
    \label{fig:spectrum_full}
\end{figure}

% *********************************************
% *********************************************
% *********************************************
\section{Stacking procedure} \label{sec:stacking}

%{\color{red} Describe RFI flagging and masking source confusion estimate via rescaling of the average.}

Throughout the paper, we adopt a standard spectral line stacking procedure \citep[see e.g.][]{Healy2019,Sinigaglia2022b}. We start by extracting \HI{} cubelets around the galaxies in our sample from the full datacubes, with aperture $(3\times$FWHM of the beam major axis, $3\times$FWHM of the beam major axis, $\pm2000$ km s$^{-1})$ in (RA, DEC, velocity). We choose these apertures to ensure that the whole flux emitted by galaxies is included in the cubelets. The angular aperture corresponds to $\sim 130$ physical kpc at $z=0.23$ (i.e. minimum aperture), larger than the typical H{\scriptsize I} disk size \citep[see e.g.,][for the $M_{\rm HI}\,-$ size relation at $z=0$]{Wang2016,Rajohnson2022}. This choice allows us to not underestimate the flux, which may happen if a smaller aperture is chosen instead, and leaves us only with the problem of subtracting the flux contamination by nearby sources. Optical coordinates and spectroscopic redshifts are used to define the center of the cubelets.

Afterwards, we integrate each cubelet over angular coordinates, obtaining a spectrum. Each spectrum at observed frequency $\nu_{\rm obs}$ is then de-redshifted to its rest-frame frequency $\nu_{\rm rf}$ through $\nu_{\rm rf}=\nu_{\rm obs}(1+z)$ and converted to units of velocity as $v=cz$. As the frequency bin width is fixed throughout the cubes ($\Delta\nu = 209$ kHz), velocity bins have different widths as a function of redshift. Therefore, to ensure that all the spectra are binned the same manner in the spectral direction, spectra are resampled to a reference spectral template, with velocity bin width $\Delta v=100\,{\rm km\,s}^{-1}$. 

We convert spectra from units of flux to units of $M_{\rm HI}$ (per velocity channel) \citep[e.g.][]{Zwaan2001}:
\begin{equation}
\centering 
    M_{\rm HI}(v) = (2.356\times10^5)\,D_{\rm L}^2\, S(v)\,(1+z)^{-1} \, {\rm M}_{\odot}\, {\rm km}^{-1} \, {\rm s} 
\end{equation}
where $D_{\rm L}$ is the luminosity distance of the galaxy in units Mpc, $S(v)$ is the 21-cm spectral flux density in units Jy and $(1+z)^{-1}$ is a correction factor accounting for the flux reduction due to the expansion of the Universe.
Lastly, we co-add all the spectra together. The stacked spectrum can then be expressed as 
\begin{equation}
    \braket{M_{\rm HI}(v)} = \frac{\sum_{i=0}^{n_{\rm gal}} M_{{\rm HI},i}(v) \times \,w_i \times f_i }{\sum_{i=0}^{n_{\rm gal}} w_i \times f_i^2}
\end{equation}
where $n_{\rm gal}$ is the number of co-added spectra, and $f_i$ and $w_i$ are the average primary beam transmission and the weight assigned to each source. This equation implements primary beam correction following the procedure detailed in \cite{Gereb2013}. Eventually, we integrate the resulting stacked spectrum over $M_{\rm HI}$, in order to obtain an average $M_{\rm HI}$ estimate. We fix the integration limits to $v_{\rm lim}\pm 500 \, {\rm km \, s^{-1}}$.

Moreover, we perform our stacking experiments assuming as target quantity for the spectra not only $M_{\rm HI}$, but also $f_{\rm HI}$, as anticipated in \S\ref{sec:sample}. For the latter case, we divide each spectrum -- given in units $M_{\rm HI}$ -- by the $M_*$ of the galaxy. In this way, from the final stacked spectrum we are able compute the average $\braket{f_{\rm HI}}=\braket{M_{\rm HI}/M_*}$.   

A customary choice for the weighting scheme is to weight each spectrum for some power of the inverse of its variance. I.e., one can weight the $i$th spectrum by $w_i=1/\sigma_i^\gamma$, where $\sigma_i$ is the rms of the $i$th spectrum. The cases $\gamma=1$ \citep{Lah2007} and $\gamma=2$ \citep{Fabello2011a} have been widely employed and tested in the literature. Some authors have proposed to also weight each spectrum by distance \citep[see e.g.][]{Delhaize2013,Hu2019}. Here we adopt the perspective of \cite{Hu2019} of leaving $\gamma$ as a free parameter (the exponent was applied to distance in the case of \cite{Hu2019}), and study the evolution of the SNR and of $M_{\rm HI}$ as a function of $\gamma$. We report the results of this test in Appendix B. It turns out that the choice $\gamma=1$ maximises the SNR in all the studied subsamples. Therefore, we choose $\gamma=1$. Since we are working with a volume-limited sample, this does not induce a selection bias due to the evolution of the $M_*$ with redshift. Furthermore, we observe that $M_{\rm HI}$ slightly decreases with increasing $\gamma$. This latter effect is due to two different facts occurring simultaneously:
\begin{itemize}
\item the root mean square (rms) per channel slowly increases towards to the lower frequency part of the covered band. As a result, weights display an anticorrelation with redshift. This can be clearly seen in the left panel of Fig.~\ref{fig:weights_mhi_z}, where we plot the weights (assuming $\gamma=1$) as a function of redshift as blue circles. The anticorrelation can be appreciated by looking at the orange dashed curve, representing the median of the weights in different redshift bins. The existing correlation between weights and redshift indicates that the larger $\gamma$ the more the weighting scheme downweights galaxies at higher redshift;
\item $M_{\rm HI}$ is found to undergo a significant evolution with redshift across the probed redshift range. To demonstrate this, we have subdivided the full redshift range into two redshift subsamples -- a low-redshift subsample at $0.23<z<0.35$ and a high-redshift subsamples at $0.35<z<0.49$. In both cases we find robust detections, the former at $\sim 6.6\sigma$, the latter at $\sim 9.9\sigma$. The resulting average \HI{} masses that we measure from the stacks are $M_{\rm HI}\sim (5.91 \pm 0.90)\times 10^9 \, {\rm M_\odot}$ and $M_{\rm HI}\sim (11.64 \pm 1.18)\times 10^9 \, {\rm M_\odot}$ for the low-z and high-z subsamples, respectively. I.e., the difference in $M_{\rm HI}$ is a factor $\sim 2$. These findings evidence a strikingly large evolution with redshift of $M_{\rm HI}$. It is hard to conclusively tell whether this difference is entirely due to actual redshift evolution, or whether some non-obvious selection or cosmic variance effects contribute to enlarge the difference. We will follow-up this aspect in future work. 
\end{itemize}

The combination of these two effects causes a downweighting of the \HI{}-rich galaxies at higher redshift when we apply a weighting scheme. %As can be seen in the fourth column of Table \ref{tab:ws} -- showing the average weighted redshift as a function of $\gamma$ for three of the different subsamples -- the decrease in the average weighted redshift with increasing $\gamma$ is quite tiny. However, the rapidly evolving $M_{\rm HI}$ with redshift makes our results sensitive to even such a small variation of redshift.

We notice that this effect probably has an irrelevant impact on low-$z$ stacking experiments, where data typically are found to have a much more stable and weakly-evolving rms per channel as a function of frequency. However, this aspect becomes important in cases where a large bandwidth with and evolving noise level is surveyed, such as the present case.    

Despite all the arguments we have laid down above, we notice that (i) the change from $\gamma=0$ (unweighted case) to $\gamma=1$ is relatively tiny, and (ii) the variation occurs coherently in all the subsamples. Therefore, while the dependence of $M_{\rm HI}$ on the weighting scheme may induce a slight variation in the overall normalization, it does not change the relative differences in $M_{\rm HI}$ (within reasonable fluctuations). This allows us to draw conclusions on the differences in $M_{\rm HI}$ between the different subsamples without worrying about the choice of the weighting scheme. 

%Throughout the paper, we assume the \cite{Fabello2011a} weighting scheme ($w_i=1/\sigma_{\rm{ rms},i}^2$). 

To evaluate the uncertainty associated with our measurements, we rely on the baseline of the spectra as follows. The 1$\sigma$ noise uncertainty (in units $M_{\rm HI}$) is evaluated by computing the rms of the off-line noisy channels $\sigma_{\rm rms}$ of the stacked spectrum, i.e. those channels outside the spectral interval integrated to compute $M_{\rm HI}$. 

To further confirm the legitimacy of our detection, we also generate a reference spectrum obtained by stacking noise spectra (one noise spectrum per galaxy) extracted at randomized positions.
The positions of the noise spectra are obtained by adding a fixed angular offset to the centre of each galaxy cubelet in a random direction and defined over the same spectral range as the corresponding galaxy cubelet. The angular offset ($100^{\prime\prime}$) is chosen to guarantee that the reference spectrum is extracted close to the galaxy spectrum, although without overlaps. Also, we double-check that the reference spectrum of each galaxy has no overlaps with other known optical galaxies, reject it and draw a new one if there is any overlap. 
%reference spectrum containing no source emission is extracted and co-added to other reference spectra. The reference spectrum is obtained from a cubelet, with center defined adding a fixed angular offset to the centre of the galaxy cubelet in a random direction, and defined over the same spectral range as the galaxy cubelet. The angular offset is conveniently chosen to guarantee that the reference spectrum is extracted close to the galaxy spectrum, although without overlaps
%To the end of assessing the statistical significance of our detections, we assume the noise rms (root mean square) of the reference spectrum $1\sigma$ uncertainty around the zero mass level. 

We compute the integrated signal-to-noise ratio of the final stacked spectrum as:
\begin{equation}\label{eq:snr}
{\rm SNR}=\sum_i^{N_{\rm ch}} \braket{S_i} /(\sigma_{\rm rms}\sqrt{N_{\rm ch}}) 
\end{equation}
where $\braket{S_i}$ is the stacked spectrum, and $N_{\rm ch}$ is the number of channels over which the integration is performed \citep[e.g.,][]{Healy2019}. %We estimate uncertainties on the stacked spectrum by applying jackknife resampling %\citep[][]{Quenouille1949,Tukey1956} 
%to the galaxy sample, eliminating one galaxy at a time. 

To mitigate the impact of RFI on our stacking results, we adopt the following procedure. We first identify frequency bands severely characterized by strong RFI by studying the variation of the rms per channel as a function of frequency. After identifying bad RFI frequency windows, we first exclude from the sample galaxies with central frequency falling within one of these frequency intervals. Eventually, because a spectrum considered as valid based on the previous criterion can still have some portions overlapping to RFI-affected channels, we mask those portions by setting the flux to zero. 

Since this is the first time (to the knowledge of the authors) that this local masking procedure is applied to RFI-affected channels, we address its impact on the results by comparing the values of the average $M_{\rm HI}$ obtained from stacking in the different subsamples before and after masking. We report the results of this test in Appendix C, Table \ref{tab:rfi_masking}. It turns out that such a masking do not induce a systematic effect in the probed subamples, but rather a stochastic deviation from the unmasked case. In addition, the magnitude of the deviation is typically tiny, with a maximum (in absolute value) of $ -9.2\%$ for the $\delta > 0$ subsample. However, as can be seen in Table \ref{tab:rfi_masking}, the SNR always increases after masking. This demonstrates that the off-line RFI masking is advantageous in terms of gain in SNR, and does not introduce systematics in the stacking procedure. We treat the random deviation introduced by the implementation of such a masking by adding a conservative $10\%$ uncertainty in quadrature to the error coming from the measurement.

We address the problem of flux contamination due to source confusion using detailed MeerKAT-like simulated datacubes, built with the same setup as MIGHTEE-\HI{} observations. In particular, we use the \cite{Obreschkow2014} flux-limited mock galaxy catalogue, based on the SKA Simulated Skies semi-analytic simulations (S$^3$-SAX), and therefore on the physical models described in \cite{Obreschkow2009a,Obreschkow2009b,Obreschkow2009c}, 
to inject galaxies with realistic \HI{} masses and clustering into a blank synthetic datacube matching the same angular and spectral size as our observations \citep[see also][]{Elson2016,Elson2019}.
Then, using the same methodology presented in \citet{Elson2016}, we decomposed the spectrum extracted for each target galaxy into contributions from the actual target, and contributions from nearby contaminating galaxies. Following this procedure, we estimate the average level of contamination for the full sample to be $\sim 10\%$. As we are explicitly investigating environmental trends, which implies considering galaxies lying in regions with different clustering properties, we need to take into account that this fact will be translated into a different confusion contribution, depending on whether a galaxy lives e.g. in a low-density or in a high-density environment. While repeating the same computation performed by \cite{Darwish2015b,Darwish2017} to define the large scale structure environments on the simulated datacubes goes beyond the scope of the paper, we handle the confusion correction by rescaling the average detected \HI{} mass of each subsample by the ratio between the mean overdensity in the regions where galaxies in the subsample  live and the mean overdensity of the full sample. I.e.:
\begin{equation} \label{eq:confusion}
    \Delta M_{\rm HI,conf,i} = \frac{\Delta M_{\rm HI,conf,tot} \times \delta_i} {\delta_{\rm tot}} \, , 
\end{equation}
where $\Delta M_{\rm HI,conf,i}$ is the final confusion contribution to be subtracted to each subsample $i$, $\delta_i$ is the average overdensity at the position of each galaxy in the $i$th subsample, $\Delta M_{\rm HI,conf,tot}$ is the confusion contribution from the full sample ($\Delta M_{\rm HI,conf,tot}\sim 10\%$, obtained from simulations as described above), and $\delta_{\rm tot}$ is the average overdensity at the position of each galaxy for the full sample. This computation is applied to all the subsamples investigated in this work (full, $\delta> 0$/$\delta \le 0$, central/satellites/isolated, field/filaments/knots), and the quantitative results are expressed in percentage and listed in Table \ref{tab:results}.

\begin{table*}
    \centering
    %\hspace{-2.3cm}
    \begin{tabular}{lcccccccccc}
    \toprule
    \toprule
      Sample  & N. gal. &   $M_{\rm HI}$ & SNR   &   Conf.  & $M_{\rm HI, corr}$ & $f_{\rm HI}$ & $\braket{M_*}$ & $M_{*,{\rm med}}$ & $\braket{z}$ & $z_{\rm med}$\\% & Plot \\
     & & $[\times 10^9\, {\rm M_\odot}]$ & & & $[\times 10^9\, {\rm M_\odot}]$ & & $[\times 10^9\, {\rm M_\odot}]$ & $[\times 10^9\, {\rm M_\odot}]$ & & \\% & \\
    \midrule
    Full & $2875$ & $9.02\pm 0.76$ & $11.8$ & $10\%$ & $8.12\pm 0.75$ & $0.55\pm 0.05$ & $32.93\pm 10.71$ & $16.41\pm 10.71$ & $0.360$ & $0.370$ \\% & Fig.~\ref{fig:spectrum_full}\\
    \midrule
    $\delta\le 0$ & $1522$ & $7.86\pm1.04$ & $7.5$ & $4\%$ & $7.54\pm 1.10$ & $0.53 \pm 0.11$ & $26.09\pm 8.55$ & $14.17\pm 8.55$  & $0.360$ & $0.371$ \\% & Fig.~\ref{fig:spectrum_overdensity}, left\\
    $\delta>0$ & $1353$ & $10.08\pm1.04$ & $9.6$ & $18\%$ & $8.27\pm 0.94$ & $0.69 \pm 0.08$ & $39.00\pm 13.51$ & $19.82\pm 13.51$ & $0.359$ & $0.368$ \\% & Fig.~\ref{fig:spectrum_overdensity}, right\\
    \midrule
    Centrals & $451$ & $5.31^*$ & $<3$ & $10\%$ & $5.31^*$ & $0.61^*$ & $35.95\pm 10.35$ & $15.86\pm 10.35$ & $0.362$ & $0.371$ \\ % & Fig.~\ref{fig:spectrum_galtype}, left\\
    Satellites & $1172$ & $13.46\pm 1.32$ & $10.2$ & $16\%$ & $11.31\pm1.22$ & $0.79 \pm 0.08$ & $32.75\pm 10.67$ & $16.19 \pm 10.67$ & $0.360$ & $0.369$ \\ % & Fig.~\ref{fig:spectrum_galtype}, centre\\
    Isolated & $1252$ & $8.37\pm 0.90$ & $9.3$ & $4\%$ & $8.04\pm 0.95$ & $0.69\pm 0.09$ & $32.01 \pm 10.87$  & $17.07\pm 10.87$ & $0.359$ & $0.370$ \\ % & Fig.~\ref{fig:spectrum_galtype}, right\\
    \midrule
    Field & $1122$ & $6.17\pm 1.21$ & $5.1$ & $5\%$ & $5.86\pm 1.27$ & $0.42\pm 0.10$ & $31.11\pm 10.31$ & $15.75 \pm 10.31$ & $0.371$ & $0.373$ \\ % & Fig.~\ref{fig:spectrum_env}, left\\
    Filaments & $1324$ & $12.77\pm 0.90$ & $14.1$ & $9\%$ & $11.62\pm 0.90$ & $0.93\pm 0.09$ & $31.11\pm 9.79$ & $15.48\pm 9.79$ & $0.354$ & $0.362$ \\ % & Fig.~\ref{fig:spectrum_env}, centre\\
    Knots & $369$ & $4.11^*$ & $<3$ & $36\%$ & $4.11^*$ & $0.39$ & $45.28\pm 15.90$ & $22.74 \pm 15.90$ & $0.348$ & $0.344$ \\ % & Fig.~ \ref{fig:spectrum_env}, right\\
    \bottomrule
    \bottomrule
    \end{tabular}
    \caption{Results from the stacking runs performed on different galaxy subsamples. The first column lists the investigated case, the second column reports the number of stacked galaxies for each case, the third column the (uncorrected, i.e. as measured from the data) resulting $M_{\rm HI}$ and its associated uncertainties ($^*$ stands for $3\sigma$ upper limit in the case of non-detection), the fourth column the SNR, the fifth column the percentage confusion correction that we apply to the mass measurement, the sixth column the corrected (by confusion and RFI-masking uncertainty) $M_{\rm HI}$ measurements and their associated uncertainties, and the seventh column the Fig.~ and panel where the related spectrum is shown.}
    \label{tab:results}
\end{table*}

\begin{figure*}
    \centering
    \includegraphics[width=\textwidth]{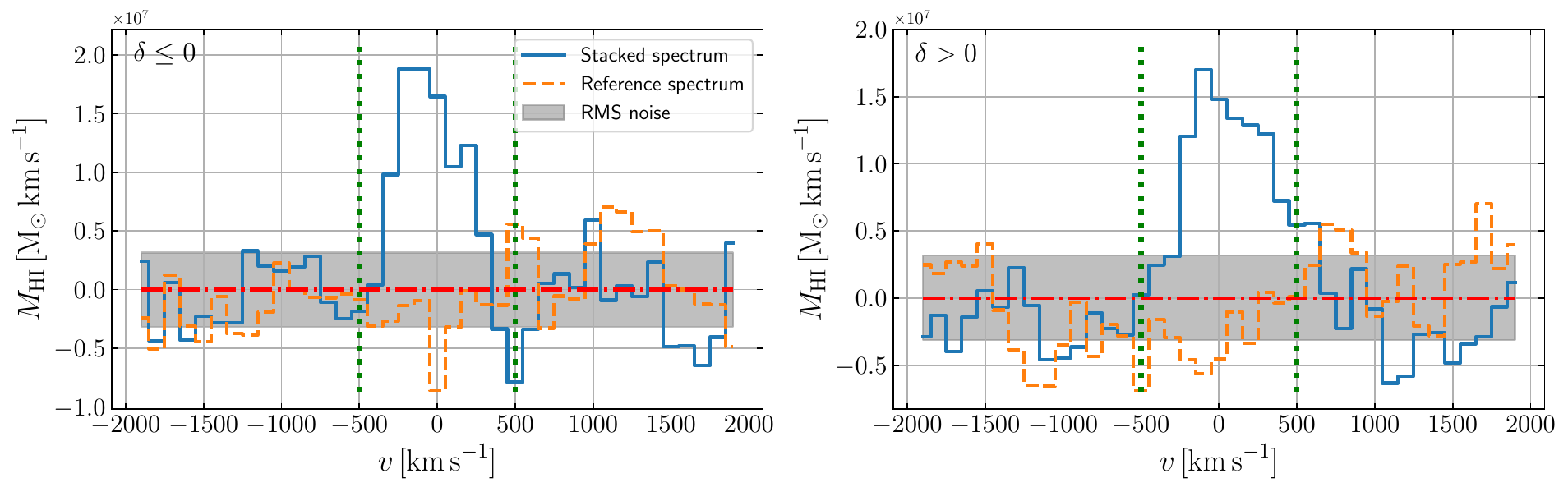}
    \caption{$M_{\rm HI}$ stacked spectrum obtained from the subsamples of galaxies in underdense ($\delta\le 0$, left) and overdense ($\delta>0$, right) regions, respectively. Symbols as in Fig.~ \ref{fig:spectrum_full}.%The blue solid line represents the stacked spectrum, the orange dashed line and the grey dashed area represent the stacked reference spectrum and its rms, respectively, and the green dotted vertical lines represent the mass integration limits.
    }
    \label{fig:spectrum_overdensity}
\end{figure*}

\begin{figure*}
    \centering
    \includegraphics[width=\textwidth]{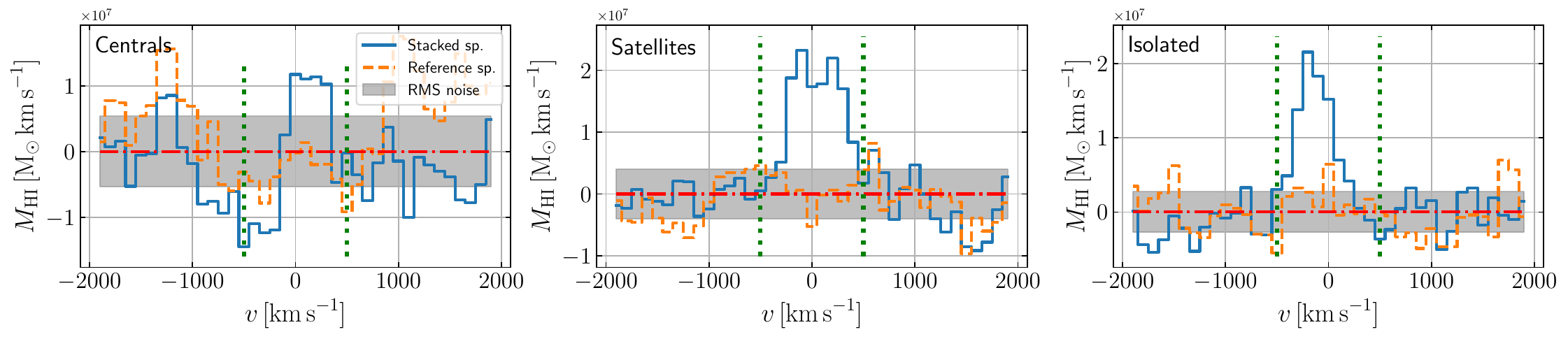}
    \caption{$M_{\rm HI}$ stacked spectrum obtained from the subsamples of central (left), satellite (centre), and isolated (right) galaxies, respectively. Symbols as in Fig.~ \ref{fig:spectrum_full}%The blue solid line represents the stacked spectrum, the orange dashed line and the grey dashed area represent the stacked reference spectrum and its rms, respectively, and the green dotted vertical lines represent the mass integration limits.
    }
    \label{fig:spectrum_galtype}
\end{figure*}

\begin{figure*}
    \centering
    \includegraphics[width=\textwidth]{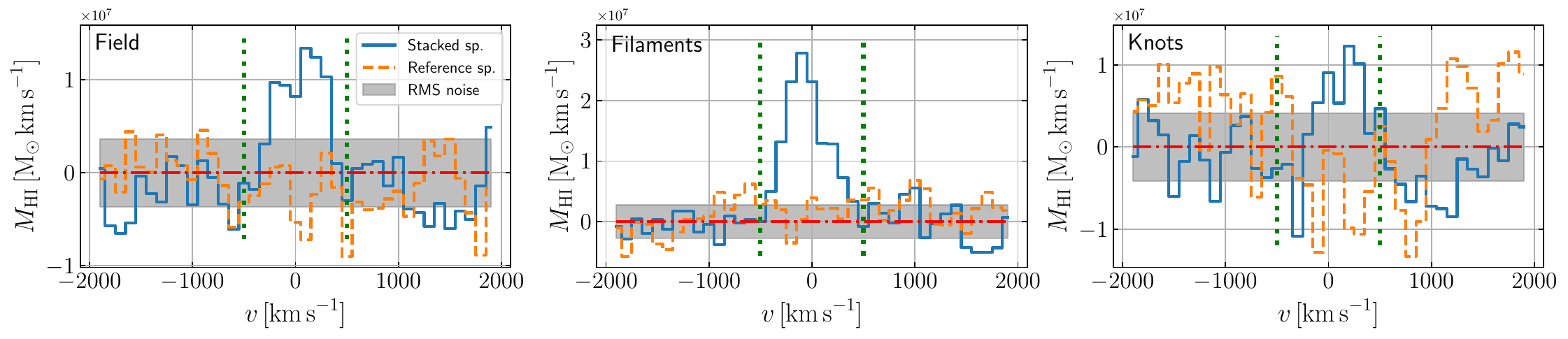}
    \caption{$M_{\rm HI}$ stacked spectrum obtained from the subsamples of galaxies sitting in the field (left), in filaments (centre), or in voids (right). Symbols as in Fig.~ \ref{fig:spectrum_full}%The blue solid line represents the stacked spectrum, the orange dashed line and the grey dashed area represent the stacked reference spectrum and its rms, respectively, and the green dotted vertical lines represent the mass integration limits.
    }
    \label{fig:spectrum_env}
\end{figure*}

\begin{figure}
    \centering
    \includegraphics[width=\columnwidth]{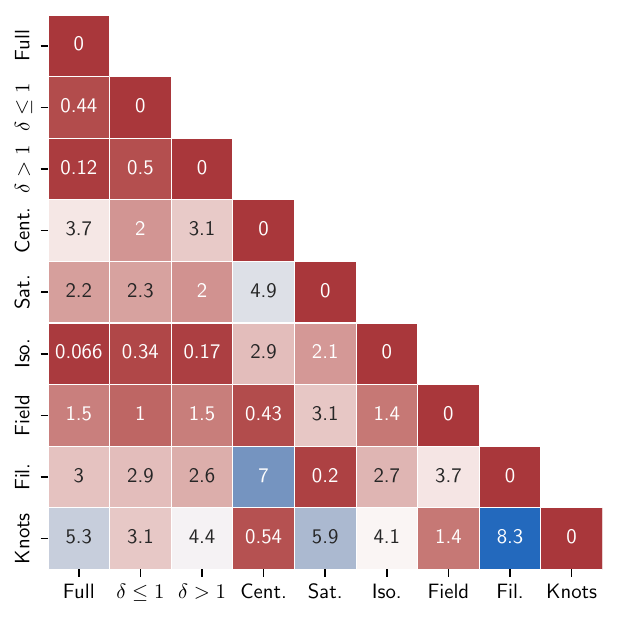}
    \caption{Statistical significance of the deviation in the average $M_{\rm HI}$ between different subsamples. The plot is color-coded from red to blue from the lowest to the highest significance values.}
    \label{fig:matrix}
\end{figure}

\begin{figure}
    \centering
    \includegraphics[width=\columnwidth]{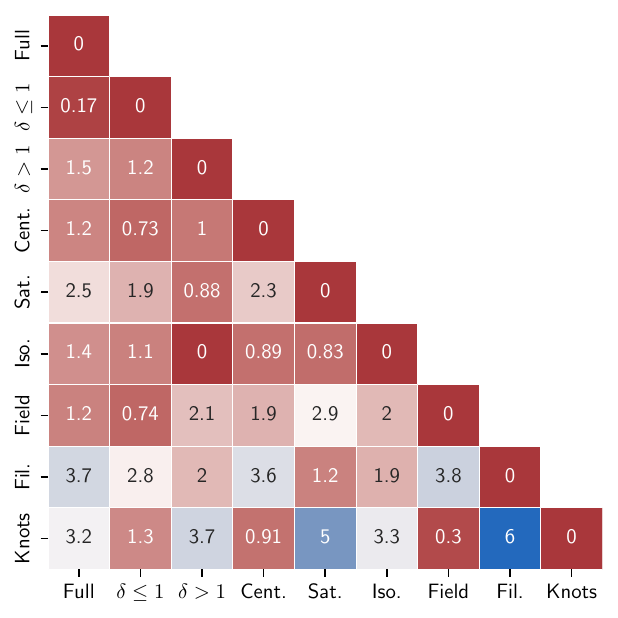}
    \caption{Statistical significance of the deviation in the average \HI{} fraction $f_{\rm HI}$ between different subsamples. The plot is color-coded from red to blue from the lowest to the highest significance values.}
    \label{fig:matrix_fraction}
\end{figure}

% *********************************************
\section{Results} \label{sec:results} 

We report on the results we obtain applying the methodology described in the previous sections, and discuss them in the light of previous findings presented in literature.

%\subsection{Results}
Figs. \ref{fig:spectrum_full} -- \ref{fig:spectrum_env}
%, Fig.~ \ref{fig:spectrum_overdensity}, Fig.~ \ref{fig:spectrum_galtype}, and Fig.~ \ref{fig:spectrum_env} 
report our resulting stacked spectra for the full sample, and for subsamples based on galaxy overdensity field, galaxy position inside the host dark matter halos, and cosmic web environment, respectively. In all cases, we display the stacked spectrum as a blue solid line, the stacked reference spectrum and its rms as an orange dashed line and a grey shaded area, respectively, and the two vertical lines marking the integration limits as green dotted lines. We report inside each panel the case each spectrum corresponds to, while we present in Table \ref{tab:results} the numerical findings, i.e. the number of stacked galaxies, the estimated average $M_{\rm HI}$ and associated uncertainty as measured from the data and after correcting for source confusion and RFI-masking uncertainty, the confusion correction that we apply, the SNR, the average and median $M_*$ of each subsample, and the average and median redshift of each subsample, the median $M_*$.%, as well as the Figure each row of the table refers to.

Overall, we find robust $>5\sigma$ \HI{} detections in all the studied case, except for the central and knot galaxies subsamples, for which we find a non-detection. The resulting stacked spectra, after applying the careful RFI flagging described in \S\ref{sec:stacking}, appear to be well-behaved, and do not display major anomalous features. %Minor negative dips are likely caused by imperfect continuum subtraction and the sidelobes of the PSF, but this is also reflected in the off-galaxy noise spectrum, hence we regard  our results to be still robust. 

In addition, we quantify the statistical significance of the deviation of the measured average $M_{\rm HI,X}$ (average $f_{\rm HI,X}$) of a subsample X from the measured average $M_{\rm HI,Y}$ (average $f_{\rm HI,X}$) of a subsample Y as:
\begin{equation}
s = \frac{|M_{\rm HI,X} - M_{\rm HI,Y}|}{\sqrt{\sigma_{\rm X}^2 + \sigma_{\rm Y}^2}} \quad ,
\end{equation}
where $\sigma_{\rm X}$ and $\sigma_{\rm Y}$ are the uncertainties on $M_{\rm HI}$ ($f_{\rm HI}$) of subsamples X and Y, respectively. We graphically report the results in Figs.~ \ref{fig:matrix} and \ref{fig:matrix_fraction}.

\begin{figure}
    \centering
    \includegraphics[width=\columnwidth]{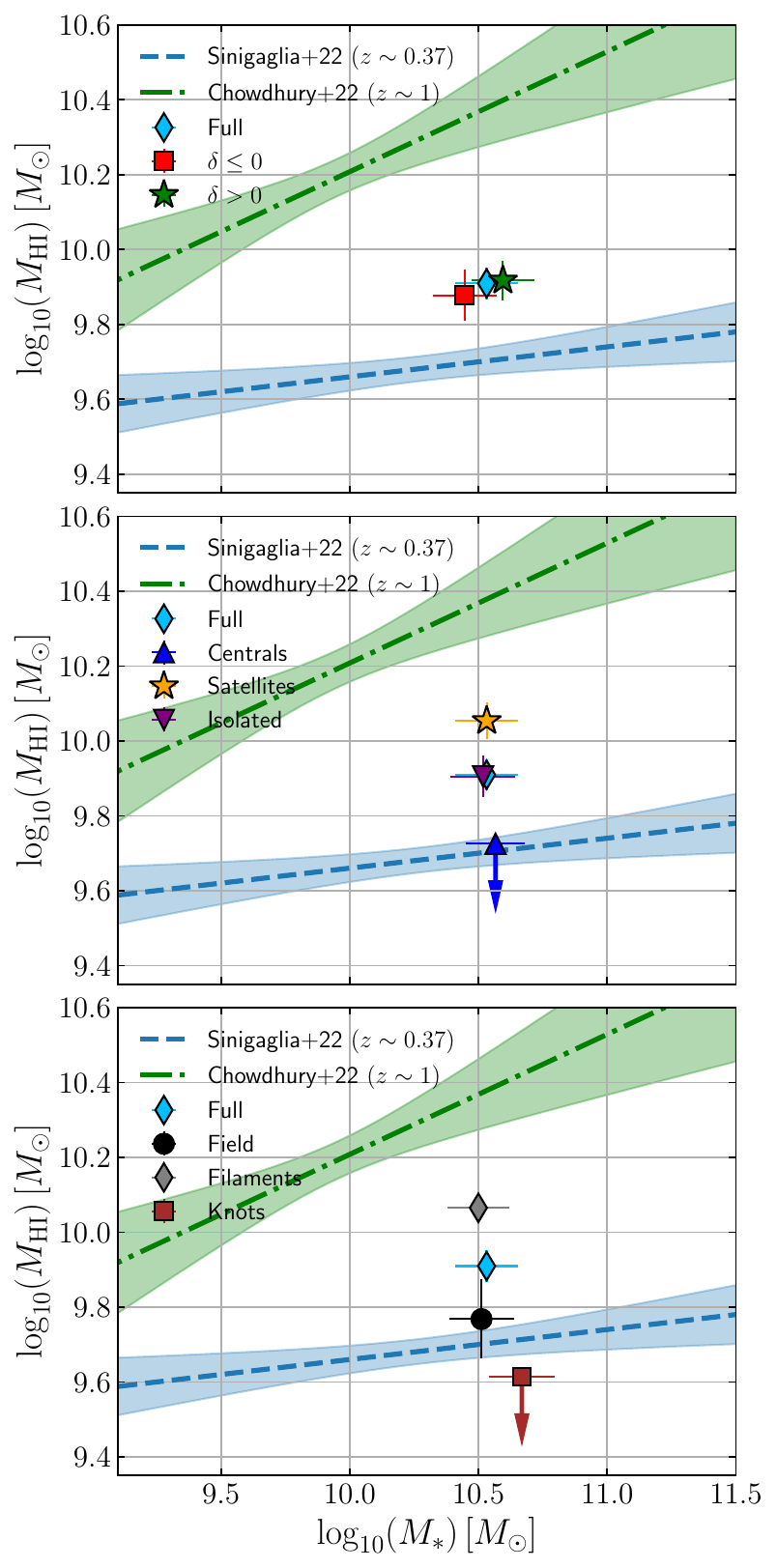}
    \caption{$M_{\rm HI}$ results as obtained from our stacking analysis in different definitions of enviroment. We compare our results to the $M_{\rm \HI{}}-M_*$ scaling relations presented in S22 at $z\sim 0.37$ and C22 at $z\sim 1$. 
    We report the findings related to different definitions of environment in different rows: galaxy overdensity (top), central/satellite/isolated (mid), field/filaments/knots (bottom). We also always overplot the data point corresponding to the full sample, for comparison purposes.}
    \label{fig:spectrum_scaling}
\end{figure}

We show the results of our stacking experiments in the $M_{\rm HI}-M_*$ plane in Fig.~\ref{fig:spectrum_scaling}. We plot the results differentiating between distinct definitions of environment: overdensity field (top), central/satellite/isolated (middle), field/filament/knot galaxies (bottom). We also plot the scaling relations at $z\sim 0.37$ from \cite{Sinigaglia2022a} (hereafter S22) and the one at $z\sim 1$ from \cite{Chowdhury2022} (hereafter C22). Since we are performing stacking, we adopt as estimate for the $M_*$ of each subsample its average $M_*$ (reported in Table \ref{tab:results}).   

We start by commenting on the results as a function of the definition of environment based on the galaxy overdensity field. Here, high- (Fig.~ \ref{fig:spectrum_overdensity}, right panel) and low-density (Fig.~ \ref{fig:spectrum_overdensity}, left panel) $M_{\rm HI}$ measurements tell that galaxies sitting in high-density environments are slightly \HI{}-richer than the ones living in low-density regions. Nonetheless, the measurements in the two subsamples are compatible within uncertainties with the $M_{\rm HI}$ estimate obtained from the full sample ($<\sim 0.5\sigma$ in both cases), and are compatible with one another (at $\sim 0.5\sigma$). The same results qualitatively apply also in terms of $f_{\rm HI}$, with larger significance ($\sim 1.5\sigma$ significance between the $\delta>0$ and the full sample, and $\sim 1.2\sigma$ between the $\delta>0$ and the $\delta\le0$ subsamples).
%However, the $M_{\rm HI}$ measurement obtained from the high-density subsample is larger by $\sim 0.1 \, {\rm dex}$ than the one obtained from the full sample, while the one obtained from low-density sample is $\sim 0.1 \, {\rm dex}$ smaller than the full sample case.

Then, we analyze the results related to the definition of environment based on galaxy position inside the host dark matter halo. It turns out that:
\begin{itemize}
\item isolated galaxies subsamples feature an average $M_{\rm HI}$ extremely similar to the one measured from the full sample %($<0.05$ dex deviation) 
and compatible within statistical uncertainties with it. Nonetheless, here again when looking at $f_{\rm HI}$ we find the isolated galaxies to be $\sim 1.4\sigma$ \HI{}-richer than the full sample;
\item central galaxies yield a non-detection both for the $M_{\rm HI}$ and the $f_{\rm HI}$ spectra. In this case we place a $3\sigma$ upper limit on $M_{\rm HI}$ and $f_{\rm HI}$. The former is found to be a factor $\sim 1.5$ \HI{} poorer than the full sample in $M_{\rm HI}$, corresponding to a $\sim 3.7\sigma$ deviation in $M_{\rm HI}$. Conversely, the upper limit for $f_{\rm HI}$ actually exceeds the \HI{} fraction of the full sample, which implies that we are not able to draw any conclusion in that sense;
\item satellite galaxies are found to be \HI{} rich, exceeding both the $M_{\rm HI}$ and the $f_{\rm HI}$ of the full sample by a factor $\sim 1.4$. This implies a deviation of $\sim 2.2\sigma$ in $M_{\rm HI}$ and $\sim 2.5\sigma$ in $f_{\rm HI}$. Moreover, satellite galaxies are \HI{}-richer than the upper limit for central galaxies by a factor $\sim 2.1$ in $M_{\rm HI}$ and a factor $\sim 1.3$ in $f_{\rm HI}$, at $\sim 4.9\sigma$ and $\sim 2.3\sigma$ respectively.
\end{itemize}
%We find this definition of environment to be the one that correlates more with the $M_{\rm HI}$ of galaxies, and hence, the one that best breaks degeneracies in the parameter space.

Finally, we examine the $M_{\rm HI}$ measurements obtained by splitting the full sample into subsamples depending on field, filaments, or knots membership. 
We find this definition of environment to be the one that correlates more with the $M_{\rm HI}$ of galaxies.
In this case: 
\begin{itemize}
\item field galaxies are found to be a factor $\sim 1.4$ \HI{}-poorer than the full sample in $M_{\rm HI}$ and a factor $\sim 1.3$ \HI{}-poorer in $f_{\rm HI}$, at $\sim 1.5\sigma$ and $\sim 1.2\sigma$, respectively;
\item filament galaxies are found to be a factor $\sim 1.4$ \HI{}-richer than the full sample in $M_{\rm HI}$ and a factor $\sim 1.7$ \HI{}-richer in $f_{\rm HI}$, at $\sim 3\sigma$ and $\sim 3.7\sigma$ respectively;
\item knots galaxies yield a non-detection. Also in this case, we place a $3\sigma$ upper limit on $M_{\rm HI}$ and $f_{\rm HI}$. Such an upper limit is a factor $\sim 2$ \HI{}-poorer than the full sample in $M_{\rm HI}$ and a factor $\sim 1.4$ \HI{}-poorer in $f_{\rm HI}$, at $\sim 5.3\sigma$ and $\sim 3.2\sigma$, respectively. Moreover, the difference between knot and filament galaxies (taking the upper limit as reference for knot galaxies) is a factor $\sim 2.8$ in $M_{\rm HI}$ and a factor $\sim 2.4$ in $f_{\rm HI}$, at $\sim 8.3\sigma$ and $\sim 6\sigma$, respectively
\end{itemize}
%While field and filaments galaxies feature a $\sim 0.1$ dex positive deviation and a $\sim 0.25$ dex negative deviation with respect to the full sample, respectively, the knots galaxies yield a very strong detection of $\log_{10}(M_{\rm HI})\sim 10.2$ and have a larger $M_{\rm HI}$ by $0.4$ dex with respect to the full sample, i.e. they appear to be a factor $\sim 2.5$ more \HI{}-rich. In fact, knot galaxies reach $M_{\rm HI}$ values in the region of the values reported by the C22 scaling relations at $z\sim 1$, being compatible within uncertainties with it.

As expected, the source confusion correction that we compute through Eq.~\S\ref{eq:confusion} is larger the denser the environment, with a minimum of $4-5\%$ for $\delta\le 0$, isolated, and field galaxies, whereas the correction is as large as $\sim 36\%$ in knots. We also notice that we do not actually apply the confusion correction to the non-detections, as it would be meaningless. 
%We stress that, despite the large confusion correction applied to the $M_{\rm HI}$ measured in knots, the resulting mass is still very large, and inconsistent with a scenario in which the excess of \HI{} can be attributed to source confusion alone.

%We draw the reader's attention to the fact that all the investigated subsamples feature similar median $M_*$, as a result of the cuts applied to low and high stellar mass to achieve completeness and to exclude red passive galaxies, respectively, and of restricting the sample only to star-forming galaxies. While this rules out the possibility of thoroughly probing different $M_*$ regimes, it offers the advantage of enabling us to perform a comparison at (nearly) fixed $M_*$.

\section{Discussion} \label{sec:discussion}

Overall, we report a tendency of intermediate-density cosmic web environments to host galaxies with larger $M_{\rm HI}$ and satellite galaxies to be more \HI{}-rich than central galaxies, the latter at (nearly) fixed stellar mass (but not at fixed halo mass). However, the $M_{\rm HI}$ correlates more with the cosmic web environments rather than with local galaxy overdensity, suggesting that environment-specific processes are in act in shaping the \HI{} content of galaxies sitting therein.  

\subsection{Central, satellite and isolated galaxies}

As anticipated, our findings at $z\sim 0.37$ suggest that satellite galaxies are \HI{}-richer than central galaxies  and isolated galaxies at fixed stellar mass $\log_{10}(M_*/{\rm M_\odot})\sim10.2$ ($\sim 2.5\sigma$).
However, it is not straightforward to interpret these results, since we are here performing a comparison at (nearly) fixed stellar mass, and not at fixed halo mass. This implies that the subsample of satellite galaxies we are investigating here are likely to be, on average, satellite of more massive central galaxies. For instance, following \cite{Shunton2022}, at $0.2<z<0.5$ central galaxies of stellar mass $\log_{10}(M_*/{\rm M}_\odot)\sim 10.2$ belong to haloes of mass $\log_{10}(M_{\rm h}/{\rm M}_\odot)\sim 12$, while satellite galaxies belong to haloes of mass $\log_{10}(M_{\rm h}/{\rm M}_\odot)\sim 12.5-13$ with characteristic central galaxy stellar mass $\log_{10}(M_*/{\rm M}_\odot)\sim 10.7-11$. Such a study was performed without splitting the sample into star-forming and passive galaxies, so both centrals and satellites in our sample of star-forming galaxies are likely to belong to haloes of lower masses with respect to the figures quoted above. However, the result remains qualitatively the same.

%These results suggest the opposite result with respect to the results at $z\sim 0$ reported in literature on central galaxies being more \HI{}-rich than satellites at the probed $M_*$. 

To put our results into context, we compare them to previous findings presented in the literature.
Using xGASS data \citep{Catinella2018} at $z\sim 0$, \cite{Janowiecki2017} find that for $\log_{10}(M_*/{\rm M_\odot})<10.2$, central galaxies in groups feature a \HI{} fraction comparable to the one of isolated galaxies at the same $M_*$ as the one probed in this work and even smaller \HI{} fraction at larger $M_*$. The same trend with $M_*$ is also mimicked by the SFR. It is important to notice that \cite{Janowiecki2017} analyze the full galaxy population (star-forming + quenched), while here we are focusing only on star-forming galaxies. By taking into account this difference and the fact that at $z\sim 0$ star-forming galaxies are known to be \HI{}-richer than passive galaxies, the results found in this work appear to agree with the ones of \cite{Janowiecki2017}. 
%According to these authors, at those redshifts, centrals living in such  small \HI{}-rich groups are found in moderately overdense environments, in an intermediate regime between isolation and clusters. The central galaxies in such groups feature a significant \HI{} reservoir that is likely fed by gas infall along filaments or from earlier minor mergers of satellites, as suggested by simulations \citep[see e.g. also][and references therein]{VillaescusaNavarro2018}. On the other hand, the central-isolated galaxies $M_{\rm HI}$ trend inverts at $\log_{10}(M_*/{\rm M_\odot})\gtrsim 10.2$ in $z\sim 0$ results by \cite{Janowiecki2017}, i.e. central galaxies become less \HI{}-massive than isolated galaxies at those high stellar masses. %To reconcile our results with such findings, we speculate that the 'transition mass' beyond which isolated galaxies become more \HI{}-rich than centrals at fixed stellar mass may shift towards higher masses going towards $z\sim 0$. 

Our results appear to also be in contrast with the combination of the observational results by \cite{Guo2021} and \cite{Brown2017} and of the simulation-based results by \cite{Stevens2019}, showing that at the fixed average $M_*$ studied in this work satellite galaxies at $z\sim 0$ are \HI{}-poorer than central galaxies \citep[see also][for a review]{Cortese2021}. %at fixed stellar mass $\log_{10}(M_*/{\rm M_\odot})\lesssim 10-10.5$.
However, again such studies investigate the full galaxy population, while we restrict our work to star-forming galaxies. Therefore, it is not possible to establish a fair comparison between those references and our results.

\subsection{Galaxies in knots, filaments and in the field}

Related to the \HI{} distribution in the cosmic web, our analysis yields a significantly larger $M_{\rm HI}$ in filaments than in knots and in the field. In this sense, we observe a significant increase in $\rm M_{\rm HI}$ from the field to filaments, i.e. from low-density to moderate/high-density environments, and then an even more significant decrease in $M_{\rm HI}$ from filaments to knots.

%These results appear to be in contrast to what is observed at $z\sim 0$. %, especially for what concerns the amount of \HI{} in large galaxy overdensities. 
While it is not trivial to establish a consistent comparison between our definition of large scale structure environment, inherited from \cite{Darwish2015b,Darwish2017}, and the distinct definitions adopted in different works at $z\sim 0$, several authors find large galaxy groups and galaxy clusters -- broadly speaking corresponding to our knots -- to be \HI{}-deficient at $z\sim 0$ \citep[e.g.][]{Giovanelli1985,Solanes2001,Gavazzi2013,Denes2014,Odekon2016,Zabel2022}.
For instance, quantifying the \HI{} deficiency as the logarithmic difference between the expected $M_{\rm HI}$ from scaling relations and the observed one \citep{Haynes1984}: 
\begin{equation}
   {\rm HI}_{\rm def} = \log_{10}(M_{\rm HI,exp}) - \log_{10}(M_{\rm HI,obs}) 
\end{equation}
\cite{Odekon2016} study the slope of the ${\rm HI}_{\rm def}-M_*$ relation for blue star-forming galaxies probed by ALFALFA \citep{Haynes2011,Haynes2018}, and find that at fixed $M_*$ the slope of such a relation rapidly increases with $M_{200}$\footnote{$M_{200}$ is the mass contained within a radius inside of which the mean interior density is $200$ times the mean density of the Universe.}. Therefore, this means that the bigger the galaxy groups, the less \HI{} is likely to be found in galaxies in such groups at fixed $M_*$.
Furthermore, focusing on satellite galaxies only, \cite{Brown2017} perform a stacking-based analysis on the ALFALFA sample and report that more massive haloes tend to host satellite galaxies with lower \HI{}-to-$M_*$ ratio, at fixed $M_*$.

The decrease in $M_{\rm HI}$ from filaments to knots indicates the presence of some gas removal or consumption mechanism. For instance, ram pressure stripping -- well-known to be a major responsible for gas removal and ionization in clusters and in particular for \HI{} stripping at $z\sim 0$ \citep[e.g.][]{Wang2020,Kleiner2023} -- may be removing \HI{} from galaxies and determine the drop in \HI{} amount from filaments to knots. In fact, ram-pressure stripping has been successfully observed to be in act in galaxy clusters up to the redshifts we are investigating \citep[$0.3<z<0.5$,][]{Moretti2022} and beyond \citep[$z\sim 0.7$,][]{Boselli2019}, suggesting that it may have been even more efficient in the past than in the present-day Universe \citep{Moretti2022}. %On the other hand, the fact that galaxies in knots are not as \HI{}-deficient as at $z\sim 0$ may have different explanations. One consists in the fact that knots and clusters are not defined the same way\footnote{We notice that \cite{Darwish2015b} uses the nomenclature 'clusters' for knots. We prefer here to use 'knots', to avoid confusion in the terminology.}. As a results, knots may contain one or more clusters, but will typically encompass a larger volume, including clusters outskirts and regions beyond clusters. Also, \HI{} could be accreted much more efficiently at $z\sim 0.37$ \citep[e.g.,][]{Sinigaglia2022a} than at $z\sim 0$ \citep[e.g.,][]{DiTeodoro2014}, and hence galaxies may still contain a significant amount of \HI{} despite the contrasting effect of other mechanisms. 

Despite the broad qualitative agreement with existing results at $z\sim 0$, we highlight three facts to be taken into account when performing such a comparison.  
First, we reiterate that we are not including passive galaxies in our study. 
Second, knots and clusters are not defined the same way\footnote{We notice that \cite{Darwish2015b} uses the nomenclature 'clusters' for knots. We prefer here to use 'knots', to avoid confusion in the terminology.}. As a results, knots may contain one or more clusters, but will typically encompass a larger volume, including clusters outskirts and regions beyond clusters.
Third, the knot galaxies subsample is the most likely to be subject to cosmic variance among the different samples investigated here. These aspects make it difficult to perform a quantitative fair comparison with $z\sim 0$ results.

Regarding the comparison between the \HI{} content in filaments and in the field, it is worth mentioning that \cite{Kleiner2017} find no significant difference between galaxies in filaments and the control sample constituted by galaxies far away from the filament spines at $z\sim 0$. However, therein filaments are defined in a different way than in this paper, hence it is again not straightfoward to perform an adequate comparison \citep[see e.g.][for a review of methods to quantitatively define the cosmic web]{Libeskind2018}. On the other hand, low-density enviroments such as cosmic voids are known to be \HI{}-poorer than filaments in cosmological hydrodynamic simulations \citep[e.g.][]{Martizzi2019}, supporting our findings.
%We will follow up this intriguing point in future papers.

Unfortunately, the limited statistics, as well as the lack of a complete spectroscopic coverage, makes it hard to perform a more detailed comparison between our definition of knots and the overdensities used in other works, as well as further investigating whether the bulk of \HI{} sits in a specific region of knots -- e.g. in the outskirt or towards the centre -- or whether it is spread across the whole radial profile. We leave this investigation for future works relying on the full area ($\sim 20$ deg$^2$) covered by the MIGHTEE survey. 

%\section{Discussion} \label{sec:discussion}

%*********************************************
%**********************************************
% *********************************************
\section{Summary and conclusions} \label{sec:conclusions}

In this work we have presented the first study of the dependence of \HI{} content in galaxies on the large scale structure, at a median redshift $z\sim 0.37$.

In particular, we select a $M_*$-complete star-forming galaxy sample ($\log_{10}(M_*/{\rm M_\odot})>9.6$) in the COSMOS field \citep[][]{Scoville2007} based on a colour-colour selection \citep[($M_{NUV}-M_r > 3(M_r - M_J) + 1, M_{NUV} - M_r > 3.1$),][]{Laigle2016} and with measured spectroscopic redshift (Khostovan et al., in prep.)

We extract individually-undetected \HI{} galaxy spectra from \HI{} datacubes delivered by the MIGHTEE survey, covering the full COSMOS field. We rely on a spectral line stacking approach to perform an average $M_{\rm HI}$ detection out of the extracted spectra for each of the studied subsample. After a careful flagging and a detailed treatment of RFI, our full sample consists of $2875$ galaxies. 

We investigate the dependence of the \HI{} content in galaxies on the large scale structure. We perform stacking on subsamples defined by restricting the parent full sample according to different definitions of the large-scale environment. In particular, we focus on the local galaxy overdensity field, the position inside the host dark matter halo (central, satellite, or isolated), and the cosmic web environment that galaxies live in (field, filaments, knots). These information are computed for each of the galaxies constituting our sample through a tomographic analysis based on photo-$z$s, and made available by \cite{Darwish2015b,Darwish2017}. We stress that the evaluation of the cosmic web environment as done in \cite{Darwish2015b}, and hence, in this work, is quantitative and based on the curvature tensor.  

Our findings can be summarized as follows:
\begin{itemize}
    \item We find a robust $\sim 11.8 \sigma$ detection of $M_{\rm HI}$ for the full sample. 
    While such a result is not fully compatible with the S22 scaling relation, the galaxy sample used in this work is not the same as in S22, and the true uncertainty associated to the scaling relation is unknown. Furthermore, the evolution of $M_{\rm HI}$ with redshift, as well as the impact of weighting scheme, makes the stacking result subject to variation depending on the reference frame one chooses. We will investigate impact of these aspects in a forthcoming paper (Bianchetti et al., in prep.)
    We point out that, because the sample is $90 \%$ complete in $M_*$ down to $\log_{10}(M_*/{\rm M_\odot})\sim 9.6$, this $M_{\rm HI}$ measurement can be exploited to compute the \HI{} density parameter $\Omega_{\rm HI}$. We leave the investigation of this aspect for future works. 
    
    %just a slight tendency, quantified as a $\sim 0.1$ dex and negative $\sim 0.05$ dex $M_{\rm HI}$ deviation from the one of the full sample, respectively, in the high- ($\delta>1$) and low-density ($\delta\le 1$) subsamples. In any case, these results are found to be compatible within the uncertainties with the $M_{\rm HI}$ measured from the parent sample.

    \item Isolated galaxies yield $M_{\rm HI}$ and $f_{\rm HI}$ measurements very similar to the one from the full sample. Central galaxies yield a non-detection and the $3\sigma$ upper limit we place corresponds to a $M_{\rm HI}$ value which exceed by a factor $\sim 1.5$ the one from the full sample ($\sim 3.7\sigma$ deviation). Satellite galaxies are found to be \HI{}-rich, exceeding both the $M_{\rm HI}$ and the $f_{\rm HI}$ of the full sample by a factor $\sim 1.4$, at $\sim 2.2\sigma$ and $\sim 2.5\sigma$ respectively. Moreover, satellite galaxies are \HI{}-richer than the upper limit for central galaxies by a factor $\sim 2.1$ in $M_{\rm HI}$ and by a factor $\sim 1.3$ in $f_{\rm HI}$, at $\sim 4.9\sigma$ and $\sim 2.3\sigma$ respectively. 
    
    %\item The \HI{} in galaxies is found to increase from the field to filaments, and to decrease from filaments to knots. Field galaxies feature a $\sim 0.2$ dex $M_{\rm HI}$ defect with respect to the full sample, filament galaxies exceed the full sample in $M_{\rm HI}$ by $\sim 0.1$ dex, and knots galaxies are found to have a \HI{} amount comparable to the one of the full sample. 
    \item Field galaxies are \HI-poorer than the full sample by a factor $\sim 1.4$ in $M_{\rm HI}$ and by a factor $\sim 1.3$ in $f_{\rm HI}$, at $\sim 1.5\sigma$ and $\sim 1.2\sigma$, respectively. Filament galaxies are \HI-richer than the full sample by a factor $\sim 1.4$ in $M_{\rm HI}$ and by a factor $\sim 1.7$ in $f_{\rm HI}$, at $\sim 3\sigma$ and $\sim 3.7\sigma$, respectively. Knot galaxies yield a non-detection and the upper limit we derive is \HI-poorer than the the full sample by a factor $\sim 2$ in $M_{\rm HI}$ and by a factor $\sim 1.4$ in $f_{\rm HI}$, at $\sim 5.3\sigma$ and $\sim 3.2\sigma$, respectively. Moreover, such an upper limit is \HI-poorer than the the filament galaxies sample by a factor $\sim 2.8$ in $M_{\rm HI}$ and by a factor $\sim 2.4$ in $f_{\rm HI}$, at $\sim 8.3\sigma$ and $\sim 6\sigma$, respectively 

    \item The interpretation of our results seem to indicate that there is a general tendency of the studied galaxy population to feature a larger $M_{\rm HI}$ towards intermediate-density cosmic web environments and the outskirts of dark matter haloes (although at fixed stellar mass, and not fixed halo mass). %The central/satellite/isolated classification of the stacked galaxies turns out to be the large scale structure parameter which is found to correlate most with the \HI{} content of galaxies at $z\sim 0.37$, effectively breaking degeneracies in the parameter space. %In particular, our findings at $z\sim 0.37$ seem to be with the results from the literature at $z\sim 0$ on the fact that central galaxies are more \HI{}-rich ($\sim 0.3$ dex) than isolated and satellite galaxies at the fixed probed stellar mass $\log_{10}(M_*/{\rm M}_\odot)\sim 10.2$, being the central galaxies likely to be hosted in low- and intermediate-mass haloes below typical masses at which the onset of AGN feedback occurs. On the other hand, the excess of \HI{} we observe in knots appear to be in contrast with results at $z\sim 0$, which systematically reports galaxies to be \HI{}-deficient in high-mass groups. We warn the reader that it is not trivial to consistently relate our findings in this regard with $z \sim 0$ results, as different definitions to identify overdensities are used.
    
\end{itemize}
    
We point out that the studied volume may still feature some cosmic variance effects, especially regarding the results related to knots. We argue that future MIGHTEE-\HI{} data beyond the Early Science dataset will allow us to significantly enlarge the probed volume at $0.23<z<0.49$, from the $\sim 1.5$ deg$^2$ of the COSMOS field used in this paper, to $\sim 20$ deg$^2$ of the final data release. 

Furthermore, we also notice once again that we are neglecting the contribution to the \HI{} budget from red passive galaxies. We leave this aspect to be investigated in future works (Rodighiero et al., in prep.).

We conclude that this work paves the way to future investigations exploring the connection between the \HI{} properties in galaxies at these redshifts, and the large scale structure environment.

\section*{Acknowledgements}

The authors wish to acknowledge the anonymous referee for the insightful comments they provided.
The MeerKAT telescope is operated by the South African Radio Astronomy Observatory, which is a facility of the National Research Foundation, an agency of the Department of Science and Innovation.
F.S. acknowledges the support of the doctoral grant funded by the University of Padova and by the Italian Ministry of Education, University and Research (MIUR)  and the financial support of the \textit{Fondazione Ing. Aldo Gini} fellowship. G.R. acknowledges the support from grant PRIN MIUR 2017 - 20173ML3WW$\char`_$001. M.V. acknowledges financial support from the Inter-University Institute for Data Intensive Astronomy (IDIA),  
a partnership of the University of Cape Town, the University of Pretoria and the University of the Western Cape, and from the South African Department of Science and Innovation's National Research Foundation under the
ISARP RADIOSKY2020 and RADIOMAP+ Joint Research Schemes (DSI-NRF Grant Numbers 113121 and 150551)
and the SRUG HIPPO Projects (DSI-NRF Grant Numbers 121291 and SRUG22031677). L.C. acknowledges support from the Australian Research Council Discovery Project and Future Fellowship funding schemes (DP210100337, FT180100066). Parts of this research were supported by the Australian Research Council Centre of Excellence for All Sky Astrophysics in 3 Dimensions
(ASTRO 3D), through project number CE170100013. I.P., F.S. and G.R. acknowledge support from INAF under the Large Grant 2022 funding scheme (project "MeerKAT and LOFAR Team up: a Unique Radio Window on Galaxy/AGN co-Evolution”). %I.H. acknowledges support from the the UK Science and Technology Facilities Council, the South African Radio Astronomy Observatory, and the Breakthrough Listen program. Breakthrough Listen is managed by the Breakthrough Initiatives, sponsored by the Breakthrough Prize Foundation. 
The authors also acknowledge the use of the ilifu cloud computing facility – \url{www.ilifu.ac.za}, a partnership between the University of Cape Town, the University of the Western Cape, the University of Stellenbosch, Sol Plaatje University, the Cape Peninsula University of Technology and the South African Radio Astronomy Observatory. The Ilifu facility is supported by contributions from the Inter-University Institute for Data Intensive Astronomy (IDIA – a partnership between the University of Cape Town, the University of Pretoria, the University of the Western Cape and the South African Radio Astronomy Observatory), the Computational Biology division at UCT and the Data Intensive Research Initiative of South Africa (DIRISA).

%%%%%%%%%%%%%%%%%%%%%%%%%%%%%%%%%%%%%%%%%%%%%%%%%%
\section*{Data Availability}

The MIGHTEE-\HI{} spectral cubes and source catalogue will be released as part of the first data release of the MIGHTEE survey.
 
%%%%%%%%%%%%%%%%%%%%%%%%%%%%%%%%%%%%%%%%%%%%%%%%%%

\bibliographystyle{mnras}
\bibliography{lit} % if your bibtex file is called example.bib

%%%%%%%%%%%%%%%%% APPENDICES %%%%%%%%%%%%%%%%%%%%%
%\begin{comment}

\section*{Appendix A: Kolmogorov-Smirnov tests} \label{sec:appendix}

We report here the results of the two-samples Kolmogorov-Smirnov (KS) tests we performed in order to further characterize the distributions of the galaxy properties displayed in Fig.~~\ref{fig:prop}. 

\begin{figure}
    \centering
    \includegraphics[width=\columnwidth]{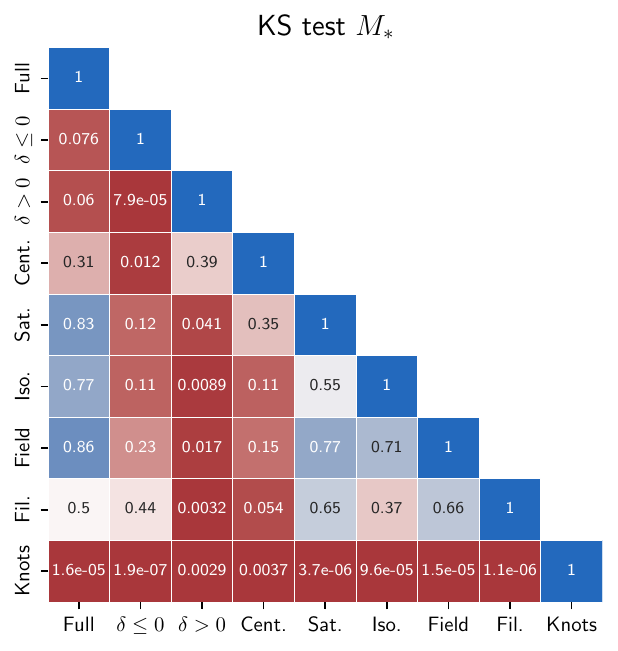}
    \caption{p-values resulting from Kolmogorov-Smirnov tests comparing $M_*$ distributions of different subsamples. The plot is color-coded from red to blue from the lowest to the highest significance values.}
    \label{fig:ks_mass}
\end{figure}

\begin{figure}
    \centering
    \includegraphics[width=\columnwidth]{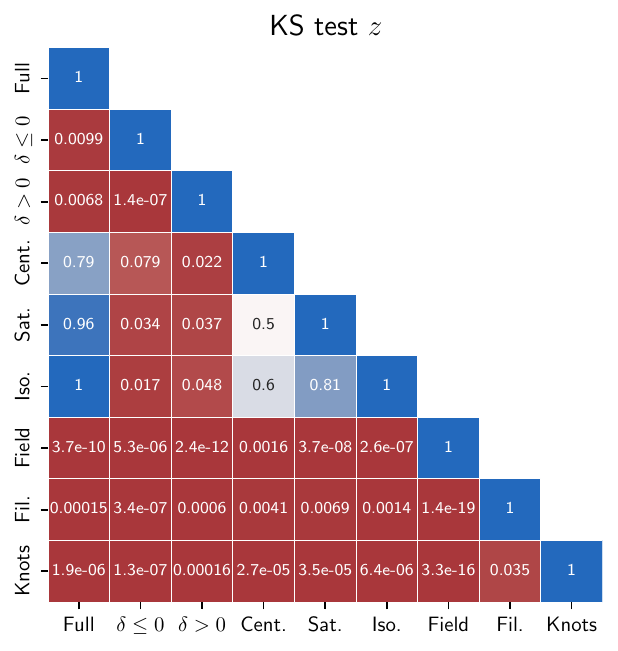}
    \caption{p-values resulting from Kolmogorov-Smirnov tests comparing $z$ distributions of different subsamples. The plot is color-coded from red to blue from the lowest to the highest significance values.}
    \label{fig:ks_z}
\end{figure}

Figs. \ref{fig:ks_mass} and \ref{fig:ks_z} show the p-values for the 1D KS tests comparing $M_*$ (Fig.~\ref{fig:ks_mass}), 
%SFR (Fig.~ \ref{fig:ks_sfr}), 
and $z$ (Fig.~\ref{fig:ks_z}) distributions for different subsamples, obtained by applying distinct cuts depending on the definition of environment. Assuming as confidence threshold $\alpha=0.05$, we reject the null hypothesis that two samples come from the same distribution if $p>\alpha=0.05$. 
In the case of the KS test comparing the $M_*$ distribution for different subsample, $\delta>0$, centrals and knots galaxies are the subsamples which tend to yield $p<0.05$. In the case of the KS test comparing the $z$ distributions, we can reject the null hypothesis in most of the cases. This means that the majority of the subsamples do not come from the same underlying distribution of $z$. 

\section*{Appendix B: Impact of the weighting scheme} \label{sec:appendix_b}

In this Section we test the impact of the weighting scheme on the average $M_{\rm HI}$ that we measure from stacking. 

\begin{figure*}
    \centering
    \includegraphics[width=\textwidth]{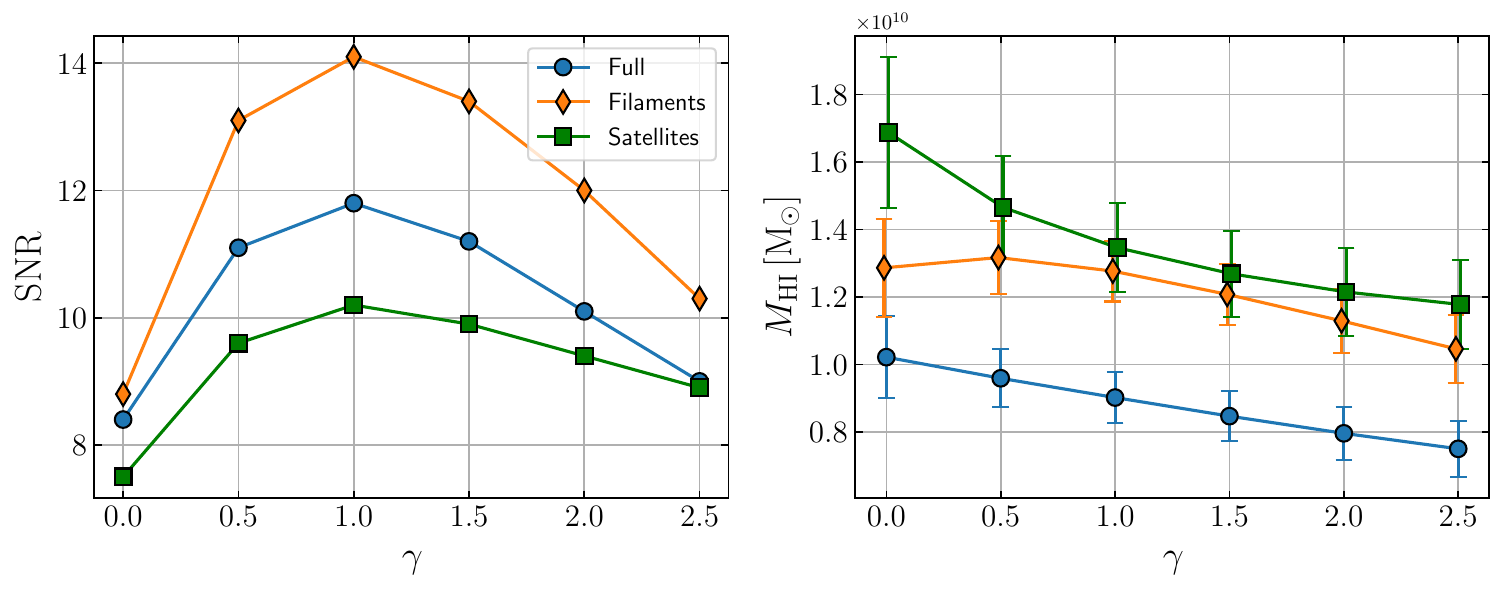}
    \caption{Left: SNR as a function of the exponent $\gamma$, for the full sample (blue circles), the filament galaxies subsample (orange diamonds), and the the satellite galaxies subsample (green squares). Right: $M_{\rm HI}$ as a function of $\gamma$, with the same colours scheme as the left panels. The same data are tabulated in Table \ref{tab:ws}.}
    \label{fig:snr_mhi_gamma}
\end{figure*}

As anticipated in \S\ref{sec:stacking}, we define the weight corresponding to the $i$th spectrum as $w_i=1/\sigma_i^\gamma$, where $\sigma_i$ is the rms of the $i$th spectrum. Subsequently, following a procedure similar in spirit to the one of \cite{Hu2019}, we study the evolution of the SNR and of $M_{\rm HI}$ as a function of $\gamma$. We report the results of the tests in Table \ref{tab:ws} and in Fig.~\ref{fig:snr_mhi_gamma}. We report the evolution of SNR (left panel) and of $M_{\rm HI}$ (right panel) as a function of $\gamma$. For the sake of clarity of visualization, we report here the results related to the full sample and the filament and satellite galaxies subsamples, and the same we do also for the numerical results in Table \ref{tab:ws}. However, we stress that we have tested all the subsamples and the same conclusions hold for all of them. It turns out that the choice $\gamma=1$ maximises the SNR in all the studied subsamples. Therefore, we choose $\gamma=1$. Since we are working with a volume-limited sample, this does not induce a selection bias due to the evolution of the $M_*$ with redshift. Furthermore, we observe that $M_{\rm HI}$ slightly decreases with increasing $\gamma$. 

\begin{figure*}
    \centering
    \includegraphics[width=\textwidth]{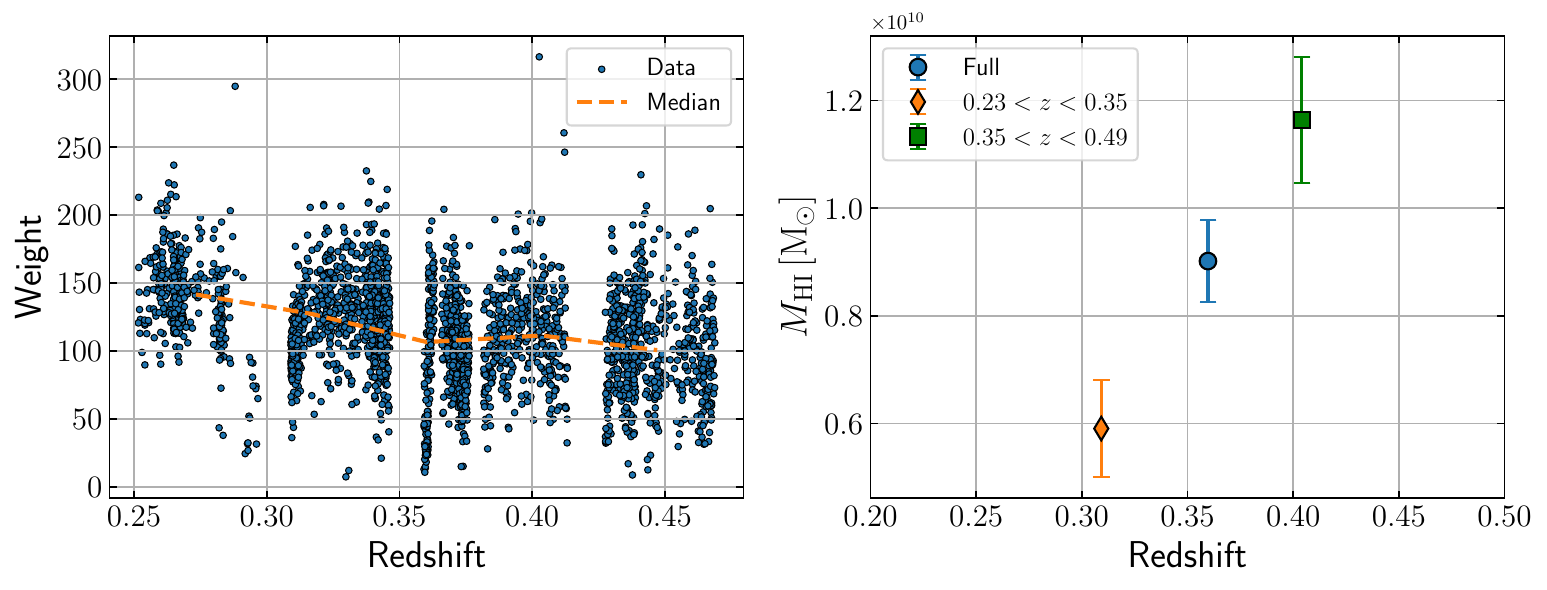}
    \caption{Left: Weights ($\gamma=1$) as a function of redshift. The median (orange dashed) evidences that there is redshift evolution in the weights distribution. Right: $M_{\rm HI}$ as a function of redshift, for the full sample (blue circle), for the subsample at $0.23<z<0.35$ (orange diamond), and for the $0.35<z<0.49$ subsample (green square).}
    \label{fig:weights_mhi_z}
\end{figure*}

To investigate the origin of the aforementioned effects, we study possible correlations between the weights (assuming $\gamma=1$) associated with the galaxies belonging to the full sample in the left panel and redshift. Also, we split the full sample into two redshift subsamples -- a low-redshift subsample at $0.23<z<0.35$ and a high-redshift subsample at $0.35<z<0.49$ -- and compare the resulting average $M_{\rm HI}$ to one we compute from the full sample. We plot the weights as a function redshift in the left panel of Fig.~\ref{fig:weights_mhi_z}, showing our data points as blue circles and the median of the weight distributions in different redshift bins as an orange dashed curve. Then, we display $M_{\rm HI}$ as a function of redshift in the right panel of Fig.~\ref{fig:weights_mhi_z}, where the blue circle stands for the full sample, the orange diamond for the low-redshift ($0.23<z<0.35$) subsample, and the green square for the high-redshift ($0.35<z<0.49$) subsample. We report an anticorrelation between weights and redshift and a positive correlation between $M_{\rm HI}$ and redshift. 

We discuss the implications of these findings in \S\ref{sec:stacking}.

\begin{table}
    \centering
    \begin{tabular}{cccc}
    \toprule
    \toprule
        & {\bf Full sample} & & \\ 
    \toprule
    $\gamma$  &  $M_{\rm HI} [\times 10^9\, {\rm M_\odot}]$ & SNR   &   $\braket{z}$ \\
    \midrule
    $0.0$ & $10.22\pm 1.21$ & $8.4$ & $0.366$ \\
    $0.5$ & $9.59\pm 0.86$ & $11.1$ & $0.363$ \\
    $1.0$ & $9.02\pm 0.76$ & $11.8$ & $0.360$ \\
    $1.5$ & $8.46\pm 0.75$ & $11.2$ & $0.357$ \\
    $2.0$ & $7.96\pm 0.79$ & $10.1$ & $0.355$ \\
    $2.5$ & $7.50\pm 0.83$ & $9.0$ & $0.353$ \\
    \bottomrule
    \bottomrule
        & {\bf Filaments} & & \\ 
    \toprule
    $\gamma$  &  $M_{\rm HI} [\times 10^9\, {\rm M_\odot}]$ & SNR   &   $\braket{z}$ \\
    \midrule
    $0.0$ & $12.86\pm 1.45$ & $8.8$ & $0.360$ \\
    $0.5$ & $13.17\pm 1.17$ & $13.1$ & $0.375$ \\
    $1.0$ & $12.77\pm 0.90$ & $14.1$ & $0.354$ \\
    $1.5$ & $12.08\pm 0.90$ & $13.4$ & $0.351$ \\
    $2.0$ & $11.29\pm 0.94$ & $12.0$ & $0.348$ \\
    $2.5$ & $10.46\pm 1.01$ & $10.3$ & $0.346$ \\
    \bottomrule
    \bottomrule
        & {\bf Satellites} & & \\ 
    \toprule
    $\gamma$  &  $M_{\rm HI} [\times 10^9\, {\rm M_\odot}]$ & SNR   &   $\braket{z}$ \\
    \midrule
    $0.0$ & $16.87\pm 2.25$ & $7.5$ & $0.366$ \\
    $0.5$ & $14.65\pm 1.53$ & $9.6$ & $0.362$ \\
    $1.0$ & $13.46\pm 1.42$& $10.2$ & $0.360$ \\
    $1.5$ & $12.68\pm 1.28$ & $9.9$ & $0.357$ \\
    $2.0$ & $12.15\pm 1.29$ & $9.4$ & $0.354$ \\
    $2.5$ & $11.77\pm 1.33$ & $8.9$ & $0.352$ \\
    \bottomrule
    \bottomrule
    
    \end{tabular}
    \caption{Results of the tests on the impact of the weighting scheme from the stacking runs performed on the full sample (top), and the filament (middle) and satellite (bottom) galaxies subsamples. The first column lists the different tested values of $\gamma$, the second column the average $M_{\rm HI}$ and its associated uncertainties measured from the data, the third column the SNR, and the fourth column the average weighted redshift. These results are graphically shown in Fig.~\ref{fig:snr_mhi_gamma}.}
    \label{tab:ws}
\end{table}

\section*{Appendix C: Impact of RFI masking} \label{sec:appendix_c}

In this section we address the effect of masking the RFI-affected channels of spectra which have been accepted in the selection procedure. This the case when the spectroscopic redshift of a galaxy -- used to define the central frequency channel of the spectrum -- does not fall within a frequency region affected by RFI, but some other channel(s) of the spectrum do. In our methodology, the flux in such channels is set to zero. In what follows, we dub this procedure 'off-line' RFI masking.

We perform the following experiment. For each subsample, we run the stacking pipeline both with and without off-line RFI masking. We report the results in Table \ref{tab:rfi_masking}. Therein, the first column lists the different investigated subsamples. The second and third columns report the average $M_{\rm HI}$ with its associated uncertainty and the SNR, respectively, before RFI masking. The third and fourth columns report the average $M_{\rm HI}$ with its associated uncertainty and the SNR, respectively, after RFI masking. The fifth column reports the (signed) percentage deviation $\Delta M_{\rm HI}=100\times (M_{\rm HI, w/\, mask}/M_{\rm HI, wo/\, mask}-1)$. The sixth, seventh, and eight columns report respectively the mean, average, and maximum percentage of channels which are masked in our procedure.

We notice that (i) the off-line RFI masking always increases the SNR with respect to the unmasked case, and (ii) such a procedure does not seem to introduce a systematic effect in the final results, and (iii) the $M_{\rm HI}$ after masking deviates $<10\%$ in al the studied cases, and $\lesssim 5\%$ in the majority of them.

This demonstrates that the off-line RFI masking is advantageous in terms of gain in SNR, and does not introduce systematics in the stacking procedure. We treat the random deviation introduced by the implementation of such a masking by adding a conservative $10\%$ uncertainty in quadrature to the error coming from the measurement. 

\begin{table*}
    \centering
    \begin{tabular}{lcccccccc}
    \toprule
    \toprule
    Sample  &  $M_{\rm HI} [\times 10^9\, {\rm M_\odot}]$ & SNR & $M_{\rm HI} [\times 10^9\, {\rm M_\odot}]$ & SNR & $\Delta M_{\rm HI}$ & Mean mask. & Med. mask. & Max. mask.\\
      & (wo/ mask) & (wo/ mask) & (w/ mask) & (w/mask) & [\%] & chan. [\%] & chan. [\%] & chan. [\%] \\
    \toprule
    Full & $8.63\pm 0.82$ & $10.5$ & $9.02 \pm 0.76$ & $11.8$ & $-4.3$ & $22.2$ & $20.2$ & $50.5$\\
    \midrule
    $\delta \le 0$ & $8.11\pm 1.20$ & $6.7$ & $7.86\pm 1.04$ & $7.5$ & $+3.2$ & $21.1$ & $19.2$ & $50.5$\\
    $\delta > 0$ & $9.15\pm 1.17$ & $8.1$ & $10.08\pm1.04$ & $9.6$ & $-9.22$ & $23.2$ & $21.2$ & $49.5$\\
    \midrule
    %Centrals & $-\pm -$ & $-$ & $\pm$ & $-$ & $-$ & $0.0$ & $0.0$ & $0.0$\\
    Satellites & $12.90\pm 1.45$ & $8.9$ & $13.46 \pm 1.32$ & $10.2$ & $-4.1$ & $22.5$ & $21.2$ & $49.5$\\
    Isolated  & $7.96\pm 1.30$ & $6.1$ & $8.37 \pm 0.90$ & $9.3$ & $-5.0$ & $22.0$ & $20.2$ & $49.5$\\
    \midrule
    Field & $6.51\pm 1.33$ & $4.9$ & $6.17 \pm 1.21$ & $5.1$ & $+5.5$ & $22.6$ & $20.2$ & $50.5$\\
    Filaments & $12.42\pm 0.97$ & $12.7$ & $12.77\pm 0.90$ & $14.1$ & $-2.3$ & $22.2$ & $21.2$ & $49.5$\\
    %Knots & $10.22\pm 1.21$ & $8.4$ & $0.366$ & $0.0$ & $0.0$ & $0.0$ & $0.0$ & $0.0$\\
    \bottomrule
    \bottomrule
    \end{tabular}
    \caption{Results of the tests on the impact of the off-line RFI masking. The first column lists the different investigated subsamples. The second and third columns report the average $M_{\rm HI}$ with its associated uncertainty and the SNR, respectively, before RFI masking. The third and fourth columns report the average $M_{\rm HI}$ with its associated uncertainty and the SNR, respectively, after RFI masking. The fifth column reports the (signed) percentage deviation $\Delta M_{\rm HI}=100\times (M_{\rm HI, w/\, mask}/M_{\rm HI, wo/\, mask}-1)$. The sixth, seventh, and eight columns report respectively the mean, average, and maximum percentage of channels which are masked in our procedure.}
    \label{tab:rfi_masking}

\end{table*}

%%%%%%%%%%%%%%%%%%%%%%%%%%%%%%%%%%%%%%%%%%%%%%%%%%

% Don't change these lines
\bsp	% typesetting comment
\label{lastpage}
\end{document}